%% file: thermorefringent_noise.tex
\documentclass[
    11pt,
    letterpaper,
    reprint,
    notitlepage,
    superscriptaddress,
    eqsecnum,
    aps,
    prx,
]{revtex4-2}

\usepackage{amsmath,amssymb,amsfonts} 
\usepackage{empheq}
\usepackage{physics}

\usepackage[normalem]{ulem}

\usepackage{microtype}

\usepackage[svgnames,dvipsnames]{xcolor}
\usepackage{graphicx}
\definecolor{NewBlue}{rgb}{0.1, 0.1, 0.7}
\definecolor{NewRed}{rgb}{0.7, 0.1, 0.1}
\usepackage[colorlinks,
    linkcolor=Maroon,
    citecolor=NewBlue,
    urlcolor=ForestGreen]{hyperref}

\usepackage[capitalise]{cleveref}

\usepackage[
    exponent-product=\cdot,
    range-phrase=\text{--},
    range-units=single,
    separate-uncertainty,
]{siunitx}[=v2]

\usepackage{multirow,bigdelim,bigstrut}
\usepackage{booktabs}

\renewcommand{\t}[1]{\mathrm{{#1}}}

\renewcommand{\hat}{\mathbf}
\renewcommand{\xi}{\zeta}

\newcommand{\eps}{\varepsilon}
\newcommand{\rmd}{\mathrm{d}}
\newcommand{\rme}{\mathrm{e}}
\newcommand{\rmi}{\mathrm{i}}

\newcommand{\tss}{\textsuperscript}

\newcommand{\avg}[1]{\left \langle {#1}\right \rangle}
\newcommand{\Var}[1]{\mathrm{Var}[{#1}]}

\DeclareMathOperator{\Ei}{Ei}

\newcommand{\LigoMIT}{LIGO Laboratory, Massachusetts Institute of Technology, Cambridge, MA 02139}
\newcommand{\MechMIT}{Department of Mechanical Engineering, Massachusetts Institute of Technology, Cambridge, MA 02139}
\newcommand{\PhysMIT}{Department of Physics, Massachusetts Institute of Technology, Cambridge, MA 02139}

\begin{document}

\title{Thermorefringent noise in crystalline optical materials}
\author{Serhii Kryhin}
\affiliation{\PhysMIT}
\author{Evan D. Hall}
\affiliation{\PhysMIT}
\affiliation{\LigoMIT}
\author{Vivishek Sudhir}
\affiliation{\LigoMIT}
\affiliation{\MechMIT}
\date{\today}

\begin{abstract}
Any material in thermal equilibrium exhibits fundamental thermodynamic fluctuations of its mechanical and optical properties.
Such thermodynamic fluctuations of length, elastic constants, and refractive index of 
amorphous materials \,---\,like dielectric mirror coatings and substrates\,---\,
limit the performance of today's most precise optical instruments.
Crystalline materials are increasingly employed in optical systems because of their reduced mechanical dissipation, which
implies a reduction of thermo-mechanical fluctuations.
However, the anisotropy of the crystalline state implies a fundamental source of thermal noise: depolarization induced by 
thermal fluctuations of its birefringence. We establish the theory of this effect, elucidate its consequences, discuss 
its relevance for precision optical experiments with crystalline materials, and hint at the conditions under which it can be evaded.
\end{abstract}

\maketitle


\section{Introduction} 

Optical media are surprisingly active even at arbitrarily low light intensity. 
Dissipation\,---\,thermal, mechanical, and optical\,---\,leads to fluctuations
in the optical fields that interact with them~\cite{Harry12}. For example, thermal dissipation in optical media
produces apparent temperature fluctuations that cause fluctuations in their refractive index and
length~\cite{Glen89,Wans92,Brag00}. The combination of such thermo-refractive and thermo-elastic noises 
--- so-called thermo-optic noise~\cite{Evans08} ---
limits the sensitivity of optical fiber strain sensors~\cite{Wans93,Fost08}, the frequency stability of fiber 
lasers~\cite{Fost07,Fost09}, and the utility of micro- and nanophotonic 
components~\cite{Gor04,AnetKip10,Baet18,Huang19,Papp20,Ham20}. On the other hand,
mechanical dissipation in optical media produces fluctuations in the material volume. Such Brownian noise 
in cavity spacers, mirror substrates, and reflective coatings~\cite{BragVyat03} limits
the stability of optical atomic clocks~\cite{NumTsub03,Numa04,Notc06,MatSter17} and the sensitivity of 
interferometric gravitational-wave detectors~\cite{Harry02,Vill10,ChalAdhi14,GrasEvans17,Gran20}.
(Noise due to optical dissipation, via the photo-thermoelastic and photo-thermorefractive mechanisms~\cite{Brag99},
have so far only been circumstantially implicated~\cite{VerKip12}.)
The common feature of these observations is the role of the amorphous character of optical materials.
 
In this context, crystalline optical materials have gained a reputation for their reduced 
Brownian noise~\cite{Cole13,ChalAdhi16}. 
The nature of thermo-optic noise in crystalline materials must be understood before
the full extent of their promise can be imagined.
However, prior theories of thermo-optic 
noise~\cite{Brag99,Brag00,LiuThorn00,Cerd01,BragVyat03,Gor08,Evans08,GurVyat10,HonChen13} focus on 
thermodynamic fluctuations in isotropic materials which do not directly apply to
crystalline materials (while prior measurements on a crystalline micro-cavity 
\cite{Alnis11} were apparently limited by thermo-refractive noise). 
Indeed, the hallmark of the crystalline state is the anisotropy of its
physical properties. In particular, both its thermal and optical responses can be anisotropic, implying
that thermodynamic fluctuations of its optical properties can be qualitatively different from those of
amorphous materials.

We show that anisotropic optical materials\,---\,exemplified by crystalline media\,---\,exhibit a
more complicated fluctuation of their apparent temperature than do isotropic materials. In turn,
this induces new types of noise in the electromagnetic field, such as the appearance of noise in
polarization modes orthogonal to the polarization of the incident mode. In general, the polarization
state of light acquires a noisy character, an effect we dub \emph{thermorefringent noise}.
Interference of such a noisy state of light with any independent reference will be imperfect, so that
thermo-refringent noise can appear as apparent amplitude and phase noise, which makes it particularly treacherous
and qualitatively different from thermo-optic noise in amorphous material (which appear as apparent phase noise only).
Indeed, the improved Brownian thermal noise performance of crystalline
coatings~\cite{Cole13} may ultimately be limited by thermorefringent noise. 



Our predictions apply equally to optical materials which can develop small anisotropies due to induced strain. 
This perspective is particularly germane to precision optical 
polarimetry~\cite{HallMa00,BielRiz09,DurRom10,BaiRob10,FleHod16}, such as for tests of QED~\cite{pvlas,RobRiz07}, 
and optical searches for physics beyond the Standard Model~\cite{alps10,ObMich18,LiuThal19}.

In addition to thermorefringent noise, the first-principles formalism we develop allows us to uncover
the possibility of thermodynamically-induced scattering into higher-order spatial modes, an effect that must also
exist in amorphous media, but has not been considered so far.

We finally introduce balanced homodyne polarimetry, a polarization sensitive variant of homodyne detection that can be used for
coherent cancellation of thermo-refringent noise in the case where thermo-refringent noise in the two orthogonal polarizations 
is strongly correlated.

The rest of the paper is organized as follows. 
In \cref{sec:mainresults} we briefly state the main results of the paper.
\cref{sec:formalism} expounds the general formalism that models the propagation of
classical electromagnetic waves through a thermally-active anisotropic medium. 
In \cref{sec:eqde} we extract stochastic equations of motion for the polarization components
of the field, which are then applied to the study of propagation through an anisotropic 
material [\cref{sec:anisotropic_transmission}],
reflection from a crystalline thin-film coating stack [\cref{sec:anisotropic_reflection}], and finally, reveal
the existence of thermodynamically-induced beam pointing noise [\cref{sec:pointing}]. 
In \cref{sec:detection} we describe the manifestation of thermorefringent noise in quantities that are typically 
measured in optical experiments. Finally, in \cref{sec:cohCan} we address the possibility of coherent cancellation
of thermo-refringent noise using balanced homodyne polarimetry.


\subsection{Summary of main results}
\label{sec:mainresults}

We establish a general formalism that describes any optical instruments affected by thermodynamic noise, in particular,
generalizing all previous treatments that neglected the polarization degree of freedom \cite{Brag99,Brag00,LiuThorn00,Cerd01,
BragVyat03,Gor08,Evans08,GurVyat10,HonChen13}. Employing this formalism, we produce concrete predictions for thermo-refringent
noise in two optical configurations directly relevant to today's most precise optical instruments --- gravitational-wave detectors,
optical atomic clocks --- and a host of precision polarimetry experiments \cite{alps10,ObMich18,LiuThal19,Obata2018,DeRocco2018,
Melissions2009,PVLAS2016,CameronAxions1993}.
These configurations are: the transmission of a plane-polarized
electric field through an anisotropic medium [\cref{sec:anisotropic_transmission}], 
and the reflection of a plane-polarized electric field from a periodic quarter-wavelength stack of 
alternating anisotropic thin-films [\cref{sec:anisotropic_reflection}].

The incident field is taken to propagate
along the $z$ direction, and plane-polarized along the $x$ direction, meeting the medium at normal
incidence.
The medium is characterized by the dielectric tensor $\eps_{ij}$, whose
variation with temperature $T$ is denoted $\eps'_{ij} = \partial \eps_{ij}/\partial T$. The medium is also
assumed thermally anisotropic, with a thermal diffusion tensor $D_{ij}$. In equilibrium, the local
temperature fluctuates with a characteristic intensity $\zeta^2 = 2k_B T^2/c_V$, where 
$c_V$ is the volumetric heat capacity at constant volume. 

Purely $x$-polarized light incident on a crystalline slab emerges with a fluctuations in its
incident polarization, and additional fluctuations in the $y$ direction. 
At ``small'' Fourier frequency $\Omega$, 
we show that the polarization fluctuations along the two directions are given by
(in terms of their power spectral densities of the fluctuations of the polarization
component $e_i$ along the $i$ direction):
\begin{equation*}
\begin{split}
    S_{e_x e_x}(\Omega) &=  - \dfrac{k^2 \zeta^2 |\eps_{xx}'|^2 \ell}{16 \pi n_x^2 \sqrt{D_{yy} D_{xx}}}  
        \ln \left( |\Omega \, \tau_+| \right)\\
    S_{e_y e_y}(\Omega) &= - \dfrac{k^2 \zeta^2 |\eps_{xy}'|^2 \ell}{16 \pi n_y^2 \sqrt{D_{yy} D_{xx}}} 
        \ln \left[ \left| D_{zz} \tau_+ (k\, \Delta n)^2 \right| \right],
\end{split}
\end{equation*}
where $\ell$ is the slab thickness, $k=\omega/c$ is the magnitude of the incident wave-vector, 
$\tau_+ = (r_0^2/2)(D_{xx}^{-1}+D_{yy}^{-1})$ is the transverse thermal diffusion timescale,
$r_0$ is the incident field's transverse spatial mode radius, and $\Delta n = n_x -n_y$ is the static
birefringence, with $n_{x,y}$ the static refractive indices in the transverse directions.
Here, ``small'' means that $\Omega \tau_+ \ll D_{zz} \tau_+ \omega^2 (\Delta n)^2 / c^2 \ll 1$; i.e.,
small compared to thermal diffusion in the transverse direction (but not small compared to diffusion in 
longitudinal direction). Note that $S_{e_y e_y}$ is frequency-independent.
In the ``intermediate'' frequency regime, i.e., $D_{zz} \tau_+ \omega^2 (\Delta n)^2 / c^2 \ll \Omega \tau_+ \ll 1$,
$S_{e_x e_x}$ is identical to the above expression, while
\begin{equation*}
\begin{split}
    S_{e_y e_y}(\Omega) &= - \dfrac{k^2 \zeta^2 |\eps_{xy}'|^2 \ell}{16 \pi n_y^2 \sqrt{D_{yy} D_{xx}}} 
        \ln \left( |\Omega \, \tau_+| \right)
\end{split} 
\end{equation*} 
falls logarithmically.
Finally, there is the ``large'' frequency regime, characterized by
$D_{zz} \tau_+ \omega^2 (\Delta n)^2 / c^2 \ll \Omega \tau_+$ and $1 \ll \Omega \tau_+$, in which
\begin{equation*}
\begin{split}
    S_{e_x e_x}(\Omega) &= \dfrac{k^2 \zeta^2 |\eps_{xx}'|^2 \ell \tau_+}{8 \pi n_x^2 r_0^2}
           \frac{1}{|\Omega \, \tau_+|^2} \\
    S_{e_y e_y}(\Omega) &= \dfrac{k^2 \zeta^2 |\eps_{xy}'|^2 \ell \tau_+}{8 \pi n_y^2 r_0^2}
       \frac{1}{|\Omega \, \tau_+|^2};
\end{split}
\end{equation*}
i.e., polarization noise falls as inverse square of the frequency. Note that the polarization fluctuations in the
two directions are always correlated, a detail that is discussed in \cref{sec:anisotropic_transmission}.

We then consider the question of thermorefringent noise from a high-reflector crystalline coating stack.
We model the coating as a periodic stack of a pair of quarter-wavelength crystalline thin-films of dielectric
tensors $\eps_{ij}^{(1,2)}$ (and approximately similar thermal properties, on a substrate that
is also thermally similar). When purely $x$-polarized light is incident on such a stack, the polarization
fluctuations of the reflected field are given by
\begin{equation*}
\begin{split}
    S_{e_x e_x}^\t{r}(\Omega) &= 
        \left|r_x \frac{\pi}{2}\frac{n_2 \eps_{xx}^{\prime (1)} + n_1 \eps_{xx}^{\prime (2)}}
                    {n_1 n_2 (n_1^2 - n_2^2)}
                    \right|^2 S_{\tilde{u}\tilde{u}}(\Omega) \\
    S_{e_y e_y}^\t{r}(\Omega) &= 
        \left|r_y \frac{\pi}{2}\frac{n_2 \eps_{xy}^{\prime (1)} + n_1 \eps_{xy}^{\prime (2)}}
                    {n_1 n_2 (n_1^2 - n_2^2)}
                    \right|^2 S_{\tilde{u}\tilde{u}}(\Omega),
\end{split}
\end{equation*}
where $r_{x,y}$ is the reflection
amplitude for either polarization, $n_i \approx n_{i\,x}\approx n_{i\,y}$ ($i=1,2$) the static refractive
index of each coating layer, and 
\begin{equation*}
    S_{\tilde{u}\tilde{u}}(\Omega) \approx \frac{\zeta^2}{\pi r_0^2 \sqrt{2 D_{zz} \Omega}}
\end{equation*}
is the approximate power spectral density of a temperature averaged over an
optically active region (in the ``high'' frequency regime, $\Omega \gg D_{zz}/r_0^2$). 
Exact expressions for the temperature fluctuations (including in other regimes), cross-correlation between
the polarizations, and the fate of the transmitted field, are all available in \cref{sec:anisotropic_reflection}.

In the limit of isotropic thermal and optical response, the above expressions for the polarization
fluctuations along the incident polarization can be related to known expressions for thermo-optic
noise in an isotropic material~\cite{Brag00,Brag99,BragVyat03,Evans08}.

\section{Theoretical model and formalism}
\label{sec:formalism}

\subsection{Thermodynamic fluctuations in an anisotropic body}

For a body in thermal equilibrium\,---\,described by the canonical ensemble\,---\,its energy fluctuates
with a variance~\cite{LanLif} $\Var{E} = k_B T^2 C_V$, where $T$ is the equilibrium temperature, and $C_V$ the 
heat capacity at constant volume. The energy fluctuations may be referred to an apparent temperature
fluctuation using the relation $\delta T = \delta E/C_V$ to give
\begin{equation}\label{eq:VarT}
 	\Var{T} = \frac{k_B T^2}{C_V}.
\end{equation}
We model the temperature fluctuation of the body as the spatial average of a local temperature field $u(\vb{r},t)$:
\begin{equation} \label{eq:mean_u}
	\delta T = \frac{1}{V}\int\limits_V u(\vb{r},t)\, d^3\vb{r},
\end{equation}
which is itself determined by a stochastic partial differential equation describing the transport of
local heat fluctuations in the body.
Assuming that heat transport in the body
is due to conduction, the local heat current $\dot{q}_i$ (along the $i^\t{th}$ direction) is due to temperature
gradients, and local temperature $u$ decreases by heat dissipation. This is modelled by
\begin{equation} \label{eq:Tsystem}
\begin{split}
	\dot{q}_i & = -\kappa_{ij} \partial_j u - \zeta_i (\vb{r},t)\\
	\dot{u} &= -\frac{\partial_i \dot{q}_i}{c_P},
\end{split}
\end{equation}
where $\kappa_{ij}$ is the anisotropic conductivity, $c_P$ is the volumetric heat capacity at constant pressure,
and $\zeta_i$ are stochastic heat currents modeling microscopic heat sources. 
Since we are interested in spatial regions larger than the typical extent of the microscopic heat sources
modeled by $\zeta_i$, and in time durations much slower than their typical fluctuation time scale, we take
that they are uncorrelated in space and time~\cite{Vliet71a}. 
However directional correlation needs to be determined separately. 
We consider this problem in \cref{sec:norm_constant} and conclude that the 
correlation of noise should take the form
\begin{equation}\label{eq:corr_q}
\begin{split}
    \avg{\zeta_i(\vb{r},t)} &= 0\\
    \avg{\zeta_i (\vb{r},t) \zeta_j (\vb{r},t)} &= \zeta^2\, D_{ij}\, \delta(\vb{r}-\vb{r}')\, \delta(t-t'),
\end{split}
\end{equation}
where $D_{ij} = \kappa_{ij}/c_P$ is the thermal diffusivity.
The intensity $\zeta^2$, determined so as to
be consistent with \cref{eq:VarT}, is (see \cref{sec:norm_constant})
\begin{equation}
    \zeta^2 = \frac{2 k_B T^2}{c_V},     
\end{equation} 
where $c_V = C_V/V$ is the volumetric heat capacity at constant volume.
Eliminating the heat current from \cref{eq:Tsystem} produces a stochastic partial differential equation for the temperature:
\begin{equation} \label{eq:diffusionT}
	(\partial_t - D_{ij}\partial_i \partial_j)u(\vb{r},t) = \eta(\vb{r},t),
\end{equation}
where $\eta = \partial_i \zeta_i$. 
Its formal solution,
\begin{equation}
	u(\vb{r},t) = \avg{u(\vb{r},t)} + \int\dd^3{\vb{r}}\dd{t} U(\vb{r}-\vb{r}',t-t') \eta(\vb{r}',t')\, ,
\end{equation}
is the sum of a homogeneous part $\avg{u}$, satisfying $(\partial_t-D_{ij}\partial_i \partial_j)\avg{u} = 0$, 
and a particular part, expressed in terms of the Green function $U$ of 
the operator $(\partial_t -D_{ij}\partial_i \partial_j)$ for appropriate boundary conditions. 
This sum is physically interpreted as the average temperature
field $\avg{u}$ perturbed by the fluctuation
\begin{equation}
	\delta u(\vb{r},t) \equiv \int\dd^3{\vb{r}}\dd{t} U(\vb{r}-\vb{r}',t-t') \eta(\vb{r}',t')\, .
\end{equation}
Note that since we expect $\avg{u}$ to be smooth, we can take $D_{ij}$ to be symmetric.

\subsection{Equations for electromagnetic field fluctuations}

Electromagnetic wave propagation through an anisotropic medium, whose internal temperature fluctuates 
as described above, is our primary concern. The predominant effect of temperature fluctuations in such 
a medium is a change in the relative dielectric tensor:
\begin{equation}\label{eq:epsilon}
	\eps_{ij} = \avg{\eps_{ij}} + \eps_{ij}' \delta u.
\end{equation}
Here, the coefficient $\eps_{ij}'$ may describe a temperature-dependent refractive index (along any direction), 
or the effect of temperature-dependent elastic strains which, via the photo-elastic effect, produces an apparent
refractive index change (see \cref{sec:materials}). 
In amorphous optical media the former (latter) leads to thermo-refractive~\cite{Brag00} 
(thermoelastic~\cite{Brag99}) noise.\footnote{Note that in principle, there is also a photo-thermo-refractive and photo-thermo-elastic effect\,---\,i.e. the
local temperature fluctuation in the medium is seeded by absorption of the local optical intensity\,---\,but these
effects are usually negligible in large optics.}
Restricting attention to electromagnetic field fluctuations due to temperature fluctuations that are
much slower than typical optical frequencies, the field is adiabatic with respect to the fluctuations in $\eps_{ij}$. 
The field then satisfies the Maxwell equations,
\begin{equation}
	\partial_i\partial_j E_j -\partial_j\partial_j E_i = 
	-\frac{\eps_{ij}}{c^2} \partial_t^2 E_j .
\end{equation}
Separating the fluctuation-free part of the field, i.e. $E_i = \avg{E_i} + \delta E_i$, inserting
\cref{eq:epsilon}, and linearizing gives the equation for the fluctuating part of the field,
\begin{equation}\label{eq:dE}
	\partial_i\partial_j \delta E_j -\partial_j\partial_j \delta E_i  
	+ \frac{\avg{\eps_{ij}}}{c^2} \partial_t^2 \delta E_j =
	-\frac{\eps_{ij}' \delta u}{c^2} \partial_t^2{} \avg{E_j}.
\end{equation}
which describes electric field fluctuations driven by local temperature fluctuations.

In the typical scenario of interest, the field, in the absence of temperature fluctuations, 
propagates along (say) the $z$ direction, in a pure polarization state, and in a well-characterized
spatial mode $f_0(x,y)$.
That is,
\begin{equation}\label{eq:avgE}
	\avg{E_i(\vb{r},t)} = \sqrt{P} e^{\rmi (k n_i z-\omega t)} f_0(x,y) \langle e_i^{(0)}\rangle;
\end{equation}
here $\langle e_i^{(0)}\rangle$ is a vector in the $(x,y)$ plane that denotes 
the mean polarization state; the spatial mode is normalized such that the integral of $E_i^* E_i$ in the transverse 
plane gives the optical power $P$, i.e. $\langle e_i^{(0)}\rangle$ is a unit vector, and $\abs{f_0}^2$ integrated to unity 
in the $xy$ plane.
We will only consider mean incident polarization $\langle e_i^{(0)}\rangle$ that is collinear with the principal crystal axes (i.e., 
the eigenvectors of $\avg{\eps_{ij}}$).
Fixing a spatial mode bases $\{f_\alpha\}_{\alpha =0,1,\ldots}$ that is orthonormal under the inner product,
\begin{equation*}
 	(f_\alpha|f_{\alpha'}) \equiv \int \dd x\, \dd y \, f_\alpha^* (x,y) f_{\alpha'}(x,y) ,
 \end{equation*} 
the effect of fluctuations in the medium can be studied by using the ansatz
\begin{equation}\label{eq:dEexpansion}
\begin{split}
	\delta E_i(\vb{r},t) = \sqrt{P} e^{\rmi (k n_i z-\omega t)} \sum_\alpha f_\alpha (x,y) \delta e_i^{(\alpha)}(z,t)
\end{split}
\end{equation}
that separates out the effect of the thermal fluctuation as a slow-in-time fluctuation of the polarization of the
same spatial mode ($\alpha=0$), and allows the possibility of scattering into other orthogonal modes ($\alpha \neq 0$).
The latter effect must also exist in amorphous media that exhibit thermo-optic noise, and 
must manifest as an apparent beam pointing noise; however the theoretical formalism~\cite{Levin98,Levin08} used to 
study thermo-optic noise does not directly illuminate this possibility since it focuses on a specific
observable a priori.

Note that the ansatz in \cref{eq:dEexpansion}, when restricted to the same spatial mode $f$ as that of the mean field,
i.e. $E_i = \sqrt{P}e^{\rmi (k n_i z - \omega t)} f_0(x,y)[\langle e_i^{(0)}\rangle  + \delta e_i^{(0)}]$, describes both a variation in 
length and angle of the polarization vector. When averaged over the ensemble of thermal fluctuations that cause these
polarization fluctuations, the ansatz represents a depolarized state of light.


\section{Depolarization from thermorefringent noise}\label{sec:depolarization}

We now turn to the study of the various manifestations of thermorefringent noise and the resulting depolarization of 
light. 
In \cref{sec:eqde} we derive the equations of motion for the polarization fluctuations, which are solved in 
\cref{sec:anisotropic_transmission} to estimate thermorefringent noise for transmission
through a bulk crystalline optic, while in \cref{sec:anisotropic_reflection} they are solved to estimate thermorefringent 
noise for reflection from a crystalline thin-film Bragg stack. 
\cref{sec:pointing} briefly addresses the question of scattering noise due to thermodynamic fluctuations.

\subsection{Equations of motion for the polarization fluctuations}\label{sec:eqde}

We begin by restricting attention to the case where thermal fluctuations lead to polarization fluctuations
of the same optical mode as the one that illuminates the medium of interest. We therefore neglect the 
terms proportional to the orthogonal modes $f_{\alpha\neq 0}$, then insert \cref{eq:dEexpansion,eq:avgE} into \cref{eq:dE}, and 
project out the components corresponding to the spatial mode of interest $f_0$. 
Details of this calculation are given in \cref{sec:EMevolution}.
The result are the coupled equations of motion for the polarization vectors of the mode of interest:
\begin{equation}\label{eq:motion_xy}
\begin{split}
	\left( \pdv{z} + \frac{n_x}{c}\pdv{t} \right) \delta e_x^{(0)} 
		&= \frac{ik \eps_{xx}'}{2n_x}(f_0^2|\delta u) \langle e_x^{(0)}\rangle \\
	\left( \pdv{z} + \frac{n_y}{c}\pdv{t} \right) \delta e_y^{(0)}
		&= \frac{ik \eps_{xy}' \rme^{\rmi (n_x-n_y)k z}}{2n_y} (f_0^2|\delta u) \langle e_x^{(0)}\rangle . 	
\end{split}
\end{equation}
where $n_i^2 = \eps_{ij}\delta_{ij}$ is the square of the refractive index along each direction, and $k = \omega/c$ is the in-vacuum wavevector.
In order to obtain these equations we assume adiabatic spatial variation of the transverse mode (with respect to the
spatial variation in the longitudinal direction)\,---\,which is the paraxial approximation, valid for
a Gaussian spatial mode\,---\,and adiabatic temporal variation of the noise (with respect to the timescale of
the optical frequency).
The right hand sides of \cref{eq:motion_xy} indicate that it is precisely the spatial intensity profile
of the optical field ($\propto f_0^2$) that samples the local temperature fluctuation field ($\delta u$); an aspect that
is tacitly assumed in the conventional treatment~\cite{Levin98,Levin08}, but which we derive here from first principles.


\subsection{Polarization noise in transmission through anisotropic medium} \label{sec:anisotropic_transmission}

We now consider a problem potentially relevant to any experiment where light has to traverse a crystalline material.
For example, the beamsplitters and input mirrors of interferometers consisting of crystalline coatings.
The main feature implicated by our theory is depolarization of the transmitted beam, which is also crucial for any 
precision polarimetry experiment \cite{LiuThal19,Obata2018,DeRocco2018,Melissions2009,PVLAS2016,CameronAxions1993}.

We consider a crystalline material of rectangular shape, with faces separated by a distance $\ell$, with normals along the $z$ direction.
The material is also assumed to be homogeneous in the sense that $\avg{\eps_{ij}}$ is constant at all spatial points.
Light is incident perpendicular to one of the faces, with its incoming polarization aligned along one of the principal axis 
of $\avg{\epsilon_{ij}}$, which we take to be linearly polarized along $x$ (without loss of generality). 
This can be done precisely because we have assumed $\avg{\epsilon_{ij}}$ is homogeneous; in fact, this also allows us to
assume that $\avg{\epsilon_{ij}}$ is diagonal.
We assume that the transverse extent of the material is infinitely large compared to the
optical spot size and the thermal diffusion length.
Therefore, each point of the crystal can be described by three coordinates $(x, y, z)$ where 
$x,y \in (-\infty,\infty)$, and, $z\in (0,\ell)$.

The equations of motion for the polarization fluctuations [\cref{eq:motion_xy}] can then be formally solved. 
Since they are first order hyperbolic partial differential 
equations, they can be solved along the characteristics defined by $z\pm ct/n$ \cite[\S 11.1]{MorFesh2}.
The solutions are:
\begin{align}
	\label{eq:ex_int} \delta e_x^{(0)} (z, t) &= i \frac{k \varepsilon_{xx}' }{2 n_x} \langle e_x^{(0)}\rangle 
		\int_0^z \; \dd{z'} F_x(z, z', t)\\
	\label{eq:ey_int} \delta e_y^{(0)} (z, t) &= i \frac{k \varepsilon_{xy}' }{2 n_y} \langle e_x^{(0)}\rangle 
		\int_0^z \; \dd{z'} e^{i(n_x - n_y)kz'} F_y(z, z', t)
\end{align}
where we define,
\begin{multline}
\label{eq:Fi}
	F_i(z, z', t) = \left( f_0^2(x,y) \middle| \delta u\bigl(x,y,z', t+\tfrac{n_i}{c}(z - z')\bigr) \right),
\end{multline}
which is the projection of the local temperature field on the optical intensity profile. 

In order to complete the solution we need the fluctuating local temperature field $\delta u$. The relevant
anisotropic heat equation [\cref{eq:diffusionT}] is augmented by open boundary conditions at the crystal faces.
We account for these boundary conditions via the method of images \cite[\S 12.1]{MorFesh2}: 
the problem with the open boundary conditions at $z=0,\ell$ 
is equivalent to the problem in all of space, but with sources placed 
periodically and symmetric under a mirror transformation around each of two faces. 
This equivalence allows us to simplify the problem by using the well-known Green's function of the heat operator 
in unbounded space (a straightforward generalization of well-known results \cite[\S 7.4]{MorFesh1}, 
\begin{equation*}
    U(\vb{r},t) = [(8\pi \abs{t})^{3/2} (\det \hat{D})^{1/2}]^{-1}
    \exp\left[ \frac{-r_i D_{ij}^{-1}r_j}{4\abs{t}} \right]
\end{equation*}
and modifying the source $\eta$ (rather than determining the Green's function for the confined slab geometry while retaining
the internal sources).
That is, identical sources are assumed at locations $\vb{r}' = \pm \vb{r} + 2 m \boldsymbol{\ell}$, where
$\vb{r}$ is a location of original source, $m \in \mathbb{Z}$, and $\boldsymbol{\ell} = (0,0,\ell)$ is an $\ell-$length
vector along the $z$ direction; this results in the modified source correlator,
\begin{equation}
	\avg{ \eta(\vb{r}, t) \eta(\vb{r}', t') } = \zeta^2 D_{ij} \partial_i \partial_j 
		 \sum_{\vb{s}(\vb{r}') \in S} \delta(\vb{r} - \vb{s}(\vb{r}')) \delta(t - t'),
\end{equation}
where $S$ contains all vectors of the type $\vb{s}(\vb{r}') = \pm \vb{r}' + 2 m \boldsymbol{\ell}$, for $m \in \mathbb{Z}$.
Using the Green's function, we can then write down the correlator of the temperature field:
\begin{equation}\label{eq:corr_T}
    \avg{ \delta u(\vb{r}, t) \delta u(\vb{r}', t') } = 
    \frac{\zeta^2}{2} \sum_{\vb{s}(\vb{r}') \in S} U(\vb{r}-\vb{s}(\vb{r}'),t-t'),
\end{equation}
which is essentially a sum of correlators of temperatures from each source point; 
here $\hat{D}$ is the matrix form of the diffusivity tensor.
To complete the formal solution of the polarization in \cref{eq:ex_int,eq:ey_int} we finally need the correlators of the source
terms $F_i$, the projection of the temperature field on the optical intensity profile.
Assuming a Gaussian transverse field profile, i.e. $f_0(x,y) =\exp [-(x^2+y^2) / (2 r_0^2)] / \sqrt{\pi r_0^2}$, we compute,
\begin{widetext}
\begin{align}\label{eq:corr_uu}
	\avg{ F_i^*(z'', z, t) F_j(z''', z', t+ \tau) }
	=  \frac{ \xi^2 \sum_{\vb{s}(\vb{r}') \in S} \exp \left[ - \frac{(z - s'_z(\vb{r}'))^2}{4 D_{zz} |\tau_{ij}|} \right]}
	{8 \pi^{3/2} \left[ D_{zz} |\tau_{ij}| ( 2 D_{xx} |\tau_{ij}| + r_0^2 ) (2 D_{yy} |\tau_{ij}| + r_0^2)\right]^{1/2}},
\end{align}
\end{widetext}
where $\tau_{ij} = \tau + n_j (z'''-z')/c - n_i (z'' - z)/c$.

Ultimately what is observed in an experiment are signals from photodetectors impinged by fields emanating from the medium.
We consider the various modes of detection more fully in \cref{sec:detection}, but the crux is that,
when the optical field incident on the medium has a large mean component $\avg{E_i}$, the observables derived from 
photodetection of the emanating field are linear in the field fluctuations $\delta E_i$. 
In particular, since in our model the thermodynamic source noise is Gaussian, and its transduction to optical field
fluctuations is linear, the statistical properties of the field
fluctuations are fully characterized by its spectral covariance matrix consisting of the elements ($i,j=x,y$),
\begin{equation} \label{eq:Sij_gen}
	S_{E_i^{(0)} E_j^{(0)}}(\Omega) \equiv \int\limits_{-\infty}^{+\infty} \dd \tau \,\avg{\delta E_i^{(0)*} (\ell, t) 
		\delta E^{(0)}_j (\ell, t+\tau)} e^{-\rmi \Omega \tau},
\end{equation}
where $\delta E_i^{(\alpha)} = ( f_\alpha | \delta E_i)$, and assume that the detector is placed immediately at
the exit of the crystalline slab (which is the position at which the transmitted beam is minimally depolarized~\cite{KorWolf05}).
Expressing the electric field fluctuation in terms of the polarization fluctuations [\cref{eq:dE}], and noting that the spatial
mode functions $\{f_\alpha\}$ are orthonormal, we have that,
\begin{equation}
	S_{E_i^{(0)} E_j^{(0)}}(\Omega) = P e^{\rmi k (n_i -n_j)\ell}\, S_{e_i^{(0)} e_j^{(0)}} (\omega + \Omega),
\end{equation}
where $S_{e_i^{(0)} e_j^{(0)}}(\omega + \Omega)$ are the elements of the spectral covariance matrix of the polarization
fluctuations, at a frequency $\Omega$ offset from the optical carrier at $\omega$. 
Thus, the statistical properties of the optical field that are
observable through photodetection are fully characterized by the covariance matrix of the polarization fluctuations
at offset frequencies around the carrier.

In principle \cref{eq:ex_int,eq:ey_int,eq:Fi,eq:corr_uu} contain the ingredients to compute the elements of this
covariance matrix exactly. Below we consider a few physically interesting cases.
We will exhibit the result for a crystal which is thicker than the characteristic temperature
diffusion length, i.e. $\ell \gg \sqrt{D_{zz}/\Omega}$, which is valid for large optics at room temperature. Effectively,
this approximation allows us to neglect fluctuating heat sources outside the interval $z\in [0,\ell]$, 
assume that outside this range the Gaussian function $f$ is zero, and so extend the integration limits to $z\in [-\infty,\infty]$ for the sources that are away from the crystal surface.
In this fashion, we arrive at (and dropping the superscript spatial-mode index),
\begin{align}
	S_{e_x e_x}(\Omega) &= \frac{\zeta^2 k^2 |\eps_{xx}'|^2  \ell}{16 \pi n_x^2 \sqrt{D_{yy} D_{xx}}} 
		I(\Omega, n_x \, \Omega) \label{eq:S_xx_prefinal}\\
	S_{e_y e_y}(\Omega) &= \frac{\zeta^2 k^2 |\eps_{xy}'|^2 \ell}{16 \pi n_y^2 \sqrt{D_{yy} D_{xx}}} 
		I(\Omega, \omega \Delta n + n_y \Omega) \label{eq:S_yy_prefinal}\\
	S_{e_x e_y }(\Omega) &= \frac{\zeta^2 k^2 \eps_{xx}'\eps_{xy}' \ell}{16 \pi n_x n_y \sqrt{D_{yy} D_{xx}}} 
		I(\Omega, \Delta n (\omega - \Omega)) \nonumber \\
	&\quad \times \frac{e^{i \omega \ell \Delta n/c} - e^{i \Omega \ell \Delta n/c}}
		{\Delta n (\Omega + \omega) \ell/c} \label{eq:S_xy_prefinal}
\end{align}
where $\Delta n = n_x - n_y$, and $I(\Omega_1, \Omega_2)$ is
\begin{equation}\label{eq:Iintegral}
	I(\Omega_1, \Omega_2) = \int\limits_0^{\infty} \dd\tau \; \frac{\cos \Omega_1 \tau \; 
    \exp \left[ - \frac{\Omega_2^2}{c^2} D_{zz} \tau \right]}{ \sqrt{( \tau + \tau_x) (\tau + \tau_y)}}.
\end{equation}
Here, $\tau_{x,y}$ are the characteristic diffusion times in the transverse direction of the optical field, 
$\tau_{x,y} = r_0^2 / 2 D_{xx,yy}$.

\begin{figure*}[t]
	\includegraphics[width=\columnwidth]{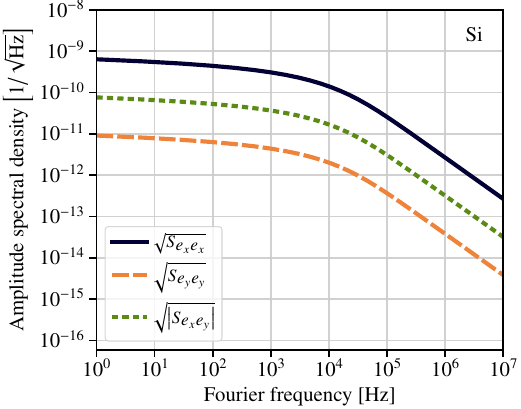}\hfill%
	\includegraphics[width=\columnwidth]{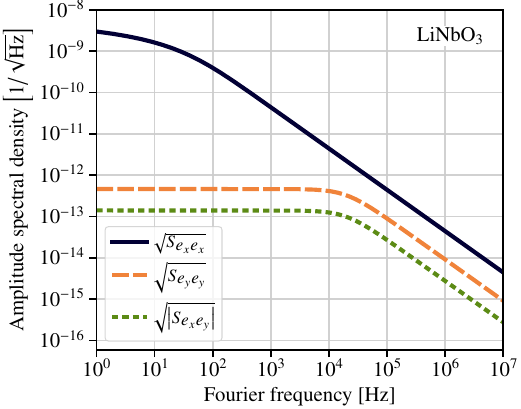}
	\caption{Thermorefringent noise for cryogenic silicon (left) and room-temperature lithium niobate (right), using \cref{eq:S_xx_prefinal,eq:S_yy_prefinal,eq:S_xy_prefinal}. In both cases the wavelength is \SI{1550}{\nm}, the beam size is $r_0 = \SI{100}{\um}$, and the material length is $\ell = \SI{1}{\cm}$.}
	\label{fig:PSD}
\end{figure*}

\Cref{fig:PSD} shows the power spectral densities \cref{eq:S_xx_prefinal,eq:S_xy_prefinal,eq:S_yy_prefinal} as applied to two different crystal systems, crystalline silicon and lithium niobate, both for a wavelength $\lambda = 2\pi c/\omega = \SI{1550}{\nm}$ and a beam size $r_0 = \SI{100}{\um}$.
The material parameters are given in \cref{tab:material_params}.
At low frequencies, below the thermal diffusion time-scale, the fluctuations in the projection of the polarization along
the direction of the incident polarization (i.e. $S_{e_x e_x }$) 
assumes a logarithmic form, turning over into a $\Omega^{-2}$ fall off. For materials for which the static birefringence is
very small ($\Delta n \ll 1$), such as crystalline silicon (\cref{fig:PSD}a), fluctuations in the other polarization, 
and the correlation between the fluctuations in either direction, also assume identical forms. 
For optical materials for which the static birefringence can be large ($\Delta n \lesssim 1$), such as lithium niobate (\cref{fig:PSD}b),
polarization fluctuations along the direction orthogonal to that of the incident polarization are strongly suppressed. 
Both types of behavior are predicted by asymptotic forms of \cref{eq:S_xx_prefinal,eq:S_yy_prefinal,eq:S_xy_prefinal}
(see \cref{sec:asymptotic}). 

It is known that if an optical standing wave is formed between the faces of a bulk amorphous medium, the resulting intensity pattern changes
the thermo-optic noise at frequencies $\Omega \sim 8k D_{zz}/r_0$ \cite{Benthem2009}. This effect is especially relevant in the input
mirrors of Fabry-Perot cavities, which cannot be wedged to avoid a standing wave in the mirror substrate.
Our formalism for the travelling wave case can be adapted to tackle the standing wave scenario. To wit, the field
\begin{equation}
	\langle \mathfrak{E}_{i} \rangle = \frac{1}{2}\left( \langle E_i(k) \rangle + \langle E_i(-k) \rangle \right)
\end{equation}
represents a standing wave $\mathfrak{E}$ as a superposition of two waves travelling in opposite directions. It then follows that
the noise in the standing wave case is 
\begin{equation}
	S_{\mathfrak{E}_i^{(0)} \mathfrak{E}_j^{(0)}}(\Omega) = \frac{1}{2} S_{E_i^{(0)} E_j^{(0)}}(\Omega) + \frac{1}{2} \t{Re}\left[ S_{E_i^{(0)}(-k) E_j^{(0)}(k)} \right].
\end{equation}
Here we have used the fact that the noise in polarization $E_i(-k)$ can be obtained from that in $E_i(k)$ by changing the sign of $k$ in
\cref{eq:ex_int,eq:ey_int}, and sign of $z-z'$ in \cref{eq:Fi}.

\input{material_params}


\subsection{Polarization noise in reflection from crystalline coating} \label{sec:anisotropic_reflection}

A standard component of contemporary precision optical instruments are low-loss mirrors composed of a stack of
dielectric thin films of alternating refractive index contrast~\cite{RemKim92}. The primary mode of 
operation of dielectric mirrors is in reflection, in which case the optical field samples a thin film stack no more than a
few tens of wavelengths deep. Despite this fact, thermal noise induced by mirror coatings dominate
precision optical instruments because these mirrors are used to recycle light within optical cavities~\cite{Harry02,Harry12}.
The nature of such thermo-optic noise in mirrors composed of amorphous dielectrics is primarily phase noise~\cite{Brag00}.
It is in this context that crystalline optical coatings were observed to be an improvement over amorphous dielectrics~\cite{Cole13}.

In this subsection we consider thermorefringent noise in a dielectric mirror made from an alternating pair of 
crystalline thin films. The two materials are described by dielectric tensors $\eps_{ij}^{(1)}, \eps_{ij}^{(2)}$,
and we assume that the mirror is made in a way that the eigenvectors of their mean dielectric tensors 
$\langle\eps_{ij}^{(1,2)}\rangle$ lie in the plane transverse to the optical axis (the latter the $z$ axis, as before).
This is true of all crystalline coatings currently being fabricated.
We further assume for simplicity that the incident light is polarized along the $x$ axis, and that the
mirror satisfies quarter-wave stack condition for this polarization.

Unlike the case of transmission through a bulk crystalline medium, the thickness of each layer in the mirror can be assumed
smaller than the thermal diffusion length of the underlying local temperature field. This is qualitatively similar to the
adiabatic limit of the transmission problem treated in \cref{sec:adiabatic}. As discussed in that context, it can be assumed
that it is the volume averaged temperature $\delta \tilde{u}$ that seeds fluctuations in the dielectric tensor. That is,
we take
\begin{align*}
	\eps_{ij}^{(I)} &= \langle\eps_{ij}^{(I)}\rangle + \eps_{ij}^{\prime (I)} \delta \tilde{u}\\ 
	&= 
	\begin{bmatrix}
		\langle[n_x^{(I)}]^2\rangle & 0 \\
		0 & \langle [n_y^{(I)}]^2 \rangle
	\end{bmatrix}
	+
	\begin{bmatrix}
		\eps_{xx}^{\prime(I)} & \eps_{xy}^{\prime(I)} \\
		\eps_{yx}^{\prime(I)} & \eps_{yy}^{\prime(I)}
	\end{bmatrix} \delta \tilde{u},
\end{align*}
where the superscript denotes the material index, $I=1,2$, and $\eps_{ij}^{\prime(I)}$ is symmetric.
The substitution of $\delta u$, which is a function of $\vb{r}$ and $t$, with some volume averaged
$\delta \tilde{u}$, which only a function of time, is an approximation valid in all the cases when fluctuating field $\delta u$ is approximately
homogeneous on the scale of the characteristic light propagation depth inside the quarter wave stack.
We give a mathematically precise definition of $\delta \tilde{u}$ later in the section.


The physical effect is that temperature fluctuations cause fluctuations in the eigenvectors of the dielectric
tensor, which is equivalent to a fluctuating birefringence.
To first order in $\delta\tilde{u}$, the above ansatz implies that the refractive indices along the two
transverse directions are
\begin{align}
	n_x^{(I)} \approx \langle n_x^{(I)}\rangle + \frac{\eps_{xx}^{\prime(I)}}{2 \langle n_x^{(I)}\rangle}\delta \tilde{u}
	\label{eq:nx}
	\\
	n_y^{(I)} \approx \langle n_y^{(I)}\rangle + \frac{\eps_{yy}^{\prime(I)}}{2 \langle n_y^{(I)}\rangle}\delta \tilde{u}
	\label{eq:ny}
\end{align}
with corresponding eigenvectors
\begin{align*}
	\vb{v}_x^{(I)} &\approx \begin{bmatrix}  1 \\ 0  \end{bmatrix} + 
		\frac{\eps_{xy}^{\prime(I)}\delta \tilde{u}}{\langle[n_x^{(I)}]\rangle^2 - \langle[n_y^{(I)}]\rangle^2}\begin{bmatrix}  0 \\ 1  \end{bmatrix}
	\\
	\vb{v}_y^{(I)} &\approx \begin{bmatrix}  0 \\ 1  \end{bmatrix} - 
		\frac{\eps_{xy}^{\prime(I)}\delta \tilde{u}}{\langle[n_x^{(I)}]\rangle^2 - \langle[n_y^{(I)}]\rangle^2}\begin{bmatrix}  1 \\ 0  \end{bmatrix}.
\end{align*}
Consequently, the eigenvectors rotate by an angle
\begin{equation}
	\delta \theta^{(I)} \approx \frac{\eps_{xy}^{\prime(I)}\delta \tilde{u}}{\langle[n_x^{(I)}]\rangle^2 - \langle[n_y^{(I)}]\rangle^2}
\end{equation}
while still retaining their length.
(Note that these expansions are valid as long as $|\eps'_{xy} \delta \tilde{u}| \ll |n_x - n_y|$.)

\begin{figure}[b]
	\includegraphics[width = 0.4 \textwidth]{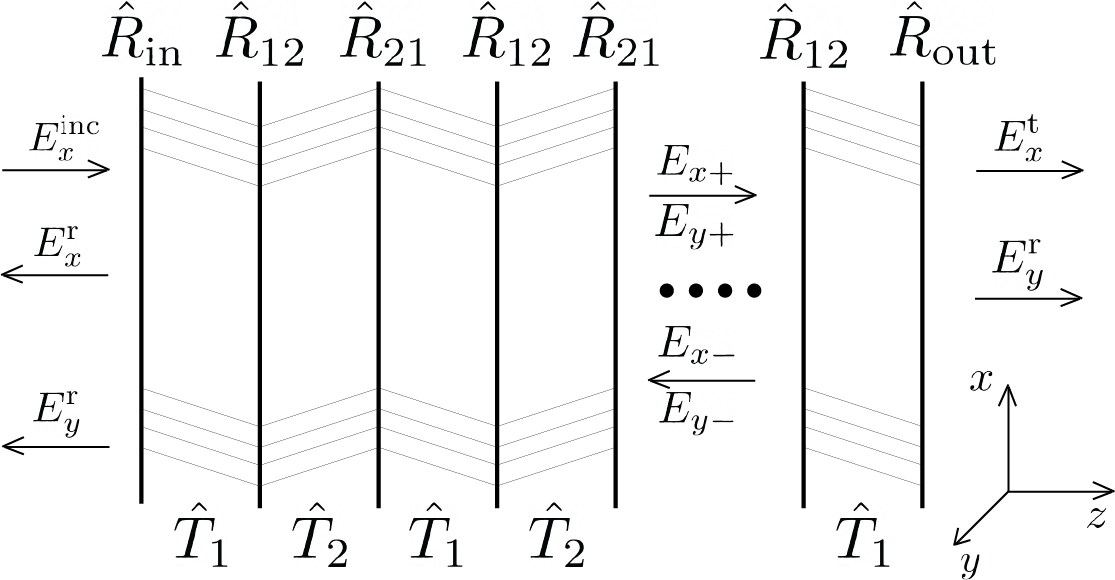}
	\centering
	\caption{Schematic picture of the mirror layers that shows the elements that
		correspond to propagation matrices and direction of relevant field modes. }
	\label{fig:mirror}
\end{figure}

\subsubsection{Transfer through a unit cell}

To study the reflection (and transmission) of the optical field from the mirror stack, we utilize the fact that
the stack is a periodic array of a simple unit cell composed of one pair of crystalline films of contrasting index,
separated by an interface at which the index jumps. This situation is illustrated in \cref{fig:mirror}. Each of the constituent films in that cell can be described by 
four fields:
\begin{equation} \label{eq:vector_components}
	\tilde{\vb{E}} = \vb{E}_x \oplus \vb{E}_y = 
  \begin{bmatrix}
    E_{x+} \\ E_{x-} \\ E_{y+} \\ E_{y-}
  \end{bmatrix}
\end{equation}
where $x$ and $y$ denote $x$ and $y$ components of the field, and
$+$ and $-$ denote propagation along positive and negative $z$ axis respectively.
We will denote the material to the left (right) in the unit cell by $I=1$ ($I=2$).
Note that the way we defined the field vector implies a definition of scattering matrices different from the common definition in optics. In our case the matrix that describes some system acts on the fields to the right of the system and returns fields to the left. The common definition acts on the vector of incident fields and returns the vector of outgoing fields. The propagation matrix that describes the passage of the field in the bulk of material $I$
can be written in the block-diagonal form
\begin{align} \label{eq:transmission_phase}
	\hat{T}^{(I)} &= \vb{T}_x^{(I)} \oplus \vb{T}_y^{(I)} = \left[
	\begin{matrix}
		\hat{T}_x^{(I)} & 0 \\
		0 & \hat{T}_y^{(I)}
	\end{matrix} \right]
	\\
	\label{eq:transmission_phase_i}
	\hat{T}_i^{(I)} &= \left[
	\begin{matrix}
		e^{-i n_i^{(I)} k z} & 0\\  
		0 & e^{i n_i^{(I)} k z}	
	\end{matrix} \right],
\end{align}
where the subscripts $x,y$ denote that the respective matrix acts on the field that is collinear
with the eigenvectors $\vb{v}_{x,y}$.
At the interface between two adjacent films, the fields are described by the
boundary conditions~\cite{LanLifEM}
\begin{align}
	\label{eq:Eboundary}
	\vb{E}^{(1)}_+ + \vb{E}^{(1)}_- = \vb{E}^{(2)}_+ + \vb{E}^{(2)}_-\\
	\label{eq:Bboundary}
	\vb{B}^{(1)}_+ + \vb{B}^{(1)}_- = \vb{B}^{(2)}_+ + \vb{B}^{(2)}_-
\end{align}
for the electric and magnetic fields.
Note that the electric and magnetic fields are related through: $c B_x  = - n_y E_y$, $c B_y = n_x E_x$. 
To write the correct matrix that describes transfer at the interface, we need to account for the relative rotation 
of the eigenvectors of the dielectric tensor between adjacent layers.
Since it is convenient to work in the basis of eigenvectors of each material, we would like to write
the interface transfer matrix in a way that it transforms the field vectors in material 2 (written in the natural basis of 
material 2) to field vectors in material 1 (written in its natural basis). 
Employing the boundary conditions in \cref{eq:Eboundary,eq:Bboundary} and accounting for the rotation of the
field vectors at the interface, we arrive at the transfer matrix
\begin{equation} \label{eq:reflection_matrix}
	\hat{R}^{(12)} = \left[
	\begin{matrix}
		\hat{r}_{xx}^{(12)} \cos \delta\theta &  -\hat{r}_{xy}^{(12)} \sin \delta\theta \\
		\hat{r}_{yx}^{(12)} \sin \delta\theta &    \hat{r}_{yy}^{(12)} \cos \delta \theta 
	\end{matrix} \right],
\end{equation}
where the matrices in each component are
\begin{equation*}
	\vb{r}_{ij}^{(IJ)} = \vb{r}(n^{(I)}_i, n^{(J)}_j); \quad i,j=x,y \quad I,J=1,2,
\end{equation*}
with $n_i^{(I)}$ being the refractive index in material $I$ along the $i$ direction,
\begin{equation} \label{eq:small_r}
	\hat{r}(a, b) \equiv \frac{1}{2} \left[
	\begin{matrix}
		1 + \frac{b}{a} & 1 - \frac{b}{a} \\
		1 - \frac{b}{a}  & 1 +\frac{b}{a}
	\end{matrix} \right],
\end{equation}
and the rotation angle is given by
\begin{equation}
	\delta \theta = \delta \theta^{(2)} - \delta \theta^{(1)}.
\end{equation}

The transfer through a single unit cell\,---\,composed of material 1 followed by material 2\,---\,is given by the
matrix
\begin{equation}\label{eq:matrix_cell}
	\hat{\Phi} =  \hat{R}^{(12)} \hat{T}^{(2)} \hat{R}^{(21)} \hat{T}^{(1)}.
\end{equation}
It describes (reading right to left), propagation through material 1, transfer at the $12$ interface, propagation
in material 2, and transfer at the $21$ interface. 
The matrix $\vb{R}^{(21)}$ is can be obtained from $\vb{R}^{(12)}$ by swapping all material indices (i.e. $1 \leftrightarrow 2$), 
and by inverting the rotation angles (i.e. $\theta \rightarrow -\theta$).

\subsubsection{Transfer through full stack}

Since the mirror is a periodic array of unit cells of the type considered above, the
transfer matrix for the stack can be expressed as a product
of the transfer matrices of each unit cell, appropriately multiplied by the transfer matrices
for the entrance coating layer and substrate interface.
Assuming that material 1 is the entrance coating, and that there are $N$ unit cells,
the transfer matrix for the mirror stack is
\begin{equation}\label{eq:matrix_mirror}
	\hat{M} = \hat{R}^{(\t{in})} \hat{T}^{(1)} \hat{\Phi}^N \hat{R}^{(\t{out})}.
\end{equation}
Here $\vb{R}^{(\t{in,out})}$ are the transfer matrices for the entrance and substrate interfaces.

The behavior of the mirror stack is encoded in the dependence of the matrix $\vb{M}$ on the polarization
angle fluctuation $\delta \theta$. It is only the factor $\vb{\Phi}^N$ that depends on the angle. 
When the angle fluctuates about zero and the fluctuations are small, we can make the expansion
\begin{equation}
    \hat{\Phi} \approx \avg{\hat{\Phi}} + \hat{\Phi}' \delta \theta,
\end{equation}
where $\avg{\vb{\Phi}}$ turns out to be block-diagonal:
\begin{align}
	\avg{\vb{\Phi}} &= \begin{bmatrix} \avg{\vb{\Phi}_{xx}} & 0\\ 0 & \avg{\vb{\Phi}_{yy}} \end{bmatrix}\\
	\label{eq:Phi_xx}
	\avg{\hat{\Phi}_{xx}} &\equiv \hat{r}_{xx}^{(12)} \hat{T}_x^{(2)} \hat{r}_{xx}^{(21)} \hat{T}_x^{(1)}\\
	\label{eq:Phi_yy}
	\avg{\hat{\Phi}_{yy}} &\equiv \hat{r}_{yy}^{(12)} \hat{T}_y^{(2)} \hat{r}_{yy}^{(21)} \hat{T}_y^{(1)},
\end{align}
and perturbation $\vb{\Phi}'$ is off-diagonal:
\begin{align}
	\hat{\Phi}' &= \left[
	\begin{matrix}
		0 & \hat{\Phi}_{xy} \\
		\hat{\Phi}_{yx} & 0
	\end{matrix} \right]
	 \\ \label{eq:Phi_xy}
	\hat{\Phi}_{xy} &= \hat{r}_{xx}^{(12)} \hat{T}_x^{(2)} \hat{r}_{xy}^{(21)} \hat{T}_y^{(1)} - 
					\hat{r}_{xy}^{(12)} \hat{T}_y^{(2)} \hat{r}_{yy}^{(21)} \hat{T}_y^{(1)}
	\\ \label{eq:Phi_yx}
	\hat{\Phi}_{yx} &= \hat{r}_{yx}^{(12)} \hat{T}_x^{(2)} \hat{r}_{xx}^{(21)} \hat{T}_x^{(1)} - 
					\hat{r}_{yy}^{(12)} \hat{T}_y^{(2)} \hat{r}_{yx}^{(21)} \hat{T}_x^{(1)}.
\end{align}
and therefore causes polarization of the reflected field to be rotated in a random manner with respect to that
of the input.

Once the mirror stack is specified, the matrices $\avg{\vb{\Phi}}$ and $\vb{\Phi}'$ can be assembled,
and the statistical properties of the resulting polarization state of the field studied. If one wants to solve 
the analogous problem for an arbitrary stack of dielectric layers, one will need to replace $\hat{\Phi}^N$ in
\cref{eq:matrix_mirror}
with $\prod_i \hat{\Phi}_i$, where $\hat \Phi_i$ is the matrix that describes the $i$\tss{th} pair of layers.

\subsubsection{Specialization to the case of a high-reflector}

Our interest here is to illustrate thermorefringent noise in a simple relevant example.
Typically, the crystalline thin film stack is configured to act as a highly reflective mirror. 
To assure the highest reflection coefficient possible, the films must satisfy a quarter-wave condition~\cite{FowlesOptics}.
This condition is typically chosen to be satisfied for one particular value of wavevector $k$, which then constrains 
the thickness of each film to be $\ell^{(I)} = \pi / (2k n^{(I)})$.
In the following we assume that this condition is met.
For typical crystalline materials used in contemporary mirrors, the in-plane optical anisotropy
is small, i.e. $\abs{\Delta n} = \abs{n_x -n_y} \ll 1$. Thus we also assume that the refractive index
along the $y$ axis is close to that along the $x$ axis: $n_y^{(I)} = n_x^{(I)} + \Delta n_I$, with
$\Delta n_I \ll 1$. In this case, the matrices $\avg{\vb{\Phi}}$ and $\vb{\Phi}'$ can be simplified by 
considering their expansions to lowest order in $\Delta n^{(I)}$. Note that according to \cref{eq:nx,eq:ny}, 
$\Delta n$ has a constant term and the term linear in $\delta u$, therefore expanding
in $\Delta n$ will reproduce an expansion in $\delta u$.
Indeed using the definitions in \cref{eq:Phi_xx,eq:Phi_yy,eq:Phi_xy,eq:Phi_yx} it can be shown that
\begin{align}
\label{eq:Phi_reflector}
	\avg{\vb{\Phi}_{xx}} &\approx \avg{\vb{\Phi}_{yy}} \approx
		-\frac{1}{2 n_1 n_2}
		\begin{bmatrix}
			n_1^2 + n_2^2  & n_1^2 - n_2^2 \\
			n_1^2 - n_2^2   & n_1^2 + n_2^2
		\end{bmatrix}\\
	\avg{\vb{\Phi}_{xy}} &= \avg{\vb{\Phi}_{yx}} \\
		&\approx \frac{\Delta n_2}{2 n_1 n_2^2}
		\begin{bmatrix}
			n_2^2 - n_1^2 - i \pi n_1 n_2 & -n_1^2 - n_2^2 \\
			-n_1^2 - n_2^2 & n_2^2 - n_1^2 + i \pi n_1 n_2
		\end{bmatrix}
\end{align}
where we have defined the direction-averaged refractive index of each film: $n_I \equiv (n_x^{(I)}+n_y^{(I)})/2$.

In order to assemble the transfer matrix $\vb{M}$ [\cref{eq:matrix_mirror}] we need a model of the entrance
coating layer (i.e. the factor $\vb{R}_\t{in} \vb{T}^{(1)}$) and the substrate (the factor $\vb{R}_\t{out}$).
The former is given by
\begin{equation} \label{eq:in_matrix}
	\hat{R}_\text{in} \hat{T}_1 
	\approx
	\begin{bmatrix}
		\hat{r}^{(1)}_x \, \hat{T}_x^{(1)} & \hat{r}^{(1)}_y \, \hat{T}_y^{(1)} \, \delta\theta_1\\
		\hat{r}^{(1)}_x \, \hat{T}_x^{(1)}\, \delta\theta_1 & \hat{r}^{(1)}_y \, \hat{T}_y^{(1)}
	\end{bmatrix},
\end{equation}
where, assuming the optical field enters from vacuum,
\begin{align*}
	\hat{r}^{(1)}_x &= \hat{r}(1, n_x^{(1)}) \\
	\hat{r}^{(1)}_y &= \hat{r}(1, n_y^{(1)}) \\
	\delta\theta^{(1)} &= \frac{\eps_{xy}^{\prime (1)} }{\avg{n_x^{(1)}}^2 - \avg{n_y^{(1)}}^2 } \delta \tilde{u}.
\end{align*}
The effect of the substrate is modelled by
\begin{equation} \label{eq:out_matrix}
	\hat{R}_\text{out} 
	\approx
	\begin{bmatrix}
		\hat{r}^s_x & -\hat{r}^s_y \, \delta\theta^{(1)}\\
		-\hat{r}^s_x \, \delta\theta^{(1)} & \hat{r}^s_y 
	\end{bmatrix},
\end{equation}
where, assuming the substrate has a refractive index $n_s$,
\begin{equation}
	\hat{r}^s_x = \hat{r}(n_s, n_{1x}), \qquad \hat{r}^s_y = \hat{r}(n_s, n_{1y}).
\end{equation}
Armed with these, the mirror matrix up to lowest order in $\delta \bar{u}$, $\Delta n_1$, and $\Delta n_2$ can be written
\begin{equation}
	\hat{M} = \avg{\hat{M}} + \vb{M}' \delta \tilde{u}.
\end{equation}
Here, $\avg{M}$ captures the static birefringence of the mirror, and is given by
\begin{equation}
	\avg{\hat{M}} = 
	\frac{- i n_1 \Gamma}{2}
	\begin{bmatrix}
		\;\;\, 1 & \;\;\, 1 & \;\;\, 0 & \;\;\, 0 \; \\
	   	       -1 &        -1 & \;\;\, 0 & \;\;\, 0 \; \\
		\;\;\, 0 & \;\;\, 0 & \;\;\, 1 & \;\;\, 1 \; \\
		\;\;\, 0 & \;\;\, 0 &        -1 &        -1 \;
	\end{bmatrix},
\end{equation}
where $\Gamma \equiv (-n_1/n_2)^N$.
The matrix $\vb{M}'$ captures the effect of thermorefringent noise, and is given by
\begin{widetext}
	\begin{equation} \label{eq:full_matrix}
		\vb{M}' = \Gamma
		\begin{bmatrix}
			- \left( \frac{i N n_1}{4} \alpha_{xx}
				+ \frac{i \eps_{xx}^{\prime (1)}}{4 n_1}  \right) \hat{A}
				- \frac{\pi \beta^+_{xx}} {8 n_2} \hat{A}_s 
			&
			\left( \frac{i N n_1}{4} \alpha_{xy}
				+ \frac{i \eps_{xy}^{\prime (1)}}{4 n_1}  
				- \frac{i N \eps_{xy}^{\prime (2)}}{2 n_2} \frac{\Delta n_2}{\Delta n_1}\right) \hat{A}
				+ \left( \frac{\beta^-_{xy}} {8 n_2}  
				+ \frac{\pi \eps_{xy}^{\prime (1)}}{4 (n_1^2 - n_2^2)} \frac{\Delta n_2}{\Delta n_1} \right) \hat{A}_s 
			\\
			- \left( \frac{i N n_1}{4} \alpha_{xy}
				+ \frac{i \eps_{xy}^{\prime (1)}}{4 n_1}  \right) \hat{A}
				- \frac{\pi \beta^+_{xy}} {8 n_2} \hat{A}_s
			&
			- \left( \frac{i N n_1}{4} \alpha_{yy}
				+ \frac{i \eps_{yy}^{\prime (1)}}{4 n_1}  \right) \hat{A}
				- \frac{\pi \beta^+_{yy}} {8 n_2} \hat{A}_s 
		\end{bmatrix}.
	\end{equation}
\end{widetext} 
Here we have defined
\begin{equation*}
\begin{split}
	\vb{A} &= \begin{bmatrix} \;\;\, 1 & \;\;\, 1 \\ -1 & -1 \end{bmatrix}, \qquad
	\vb{A}_s = \begin{bmatrix} 1 + n_s & 1 - n_s \\ 1 + n_s & 1 - n_s \end{bmatrix},\\
	\alpha_{ij} &= \left( \frac{\eps_{ij}^{\prime (1)}}{n_1^2} - \frac{\eps_{ij}^{\prime (2)}}{n_2^2} \right),\\
	\beta_{ij}^\pm &= \frac{n_2 \eps_{ij}^{\prime (1)} \pm n_1 \eps_{ij}^{\prime (2)}} {n_1^2 - n_2^2}.
\end{split}
\end{equation*}

Ultimately, we are interested in the optical fields transmitted through and reflected from the mirror stack.
When the mirror matrix $\hat{M}$ is computed, the relation
between the light in front of the mirror and behind the mirror
is given by the equation
\begin{equation} \label{eq:mirror_equation}
	\begin{bmatrix}
		E_x^\text{inc} \\ E_x^\text{r} \\ 0 \\ E_y^\text{r}
	\end{bmatrix}
	=
	\begin{bmatrix}
		M_{11} & M_{12} & M_{13} & M_{14}\\
		M_{21} & M_{22} & M_{23} & M_{24}\\
		M_{31} & M_{32} & M_{33} & M_{34}\\
		M_{41} & M_{42} & M_{43} & M_{44}
	\end{bmatrix}
	\begin{bmatrix}
		E_x^\text{t} \\ 0 \\ E_y^\text{t} \\ 0 
	\end{bmatrix},
\end{equation}
where $E_x^\text{inc}$ is an incident $x$-polarized field,
$E_{x,y}^\text{r}$ are the two polarizations of the reflected field,
and $E_{x,y}^\t{t}$ are the transmitted fields. 
In order to arrive at the conventional scattering description that relates the input
fields ($E_x^\t{inc}$) to the output fields ($E_{x,y}^\t{r,t}$), the matrix $\vb{M}$ needs
to be permuted so as to solve the linear equations \ref{eq:mirror_equation}.
Doing so gives the transmission and reflection coefficients of the high reflector stack, 
\begin{align*}
	t_x &= \frac{M_{33}}{M_{11} M_{33} - M_{13} M_{31}} \\
	t_y &= -\frac{M_{31}}{M_{11} M_{33} - M_{13} M_{31}}\\
	r_x &= \frac{M_{21} M_{33} - M_{23} M_{31}}{M_{11} M_{33} - M_{13} M_{31}} \\
	r_y &= \frac{M_{41} M_{33} - M_{43} M_{31}}{M_{11} M_{33} - M_{13} M_{31}}.
\end{align*}
Notice that these coefficients are stochastic through their dependence on $\Delta n$ (which depends
on the temperature fluctuation $\delta\tilde{u}$).
Although this dependence is nonlinear, when the fluctuations are small, in the sense that the fractional
change in the matrix element $M_{ij}$ due to $\delta \tilde{u}$, 
$\avg{M_{ij}^{-1}} (\partial M_{ij}/\partial \tilde{u}) \delta \tilde{u} \ll 1$, we can approximate the effect
of the fluctuating temperature via a linear expansion in $\delta \tilde{u}$, even for the fields.
In this fashion, we derive the fluctuating parts of the transmitted and reflected fields,
\begin{equation}\label{eq:deltaExyrt}
\begin{split}
	\delta E_x^\text{t} = - \avg{E_x^\text{t}} \frac{M_{11}'}{\avg{M_{11}}} \delta \tilde{u} \\
	\delta E_y^\text{t} = - \avg{E_x^\text{t}} \frac{M_{31}'}{\avg{M_{33}}} \delta \tilde{u} \\
	\delta E_x^\text{r} = \avg{E_x^\text{r}} 
		\left(\frac{M_{21}'}{\avg{M_{21}}} - \frac{M_{11}'}{\avg{M_{11}}} \right) \delta \tilde{u}\\
	\delta E_y^\text{r} = \avg{E_x^\text{r}} 
		\left(\frac{M_{41}'}{\avg{M_{21}}} - \frac{\avg{M_{43}} M_{31}'}{\avg{M_{33}} \avg{M_{21}}} \right) \delta \tilde{u}
\end{split}
\end{equation}
Notice that the polarization fluctuations in both transverse directions
is proportional to fluctuations in the average temperature fluctuation $\delta \tilde{u}$ in the crystalline thin-film 
stack. The reason that the volume-averaged temperature makes an appearance here, instead of a field-weighted spatial integral
of the local temperature $\delta u$ (as in \cref{sec:anisotropic_transmission}) is because the temperature field is
spatially correlated across the stack layers in the volume sampled by the optical field.

The volume-averaged temperature fluctuation $\delta \tilde{u}$ is given by a straightforward extension of 
standard results for the mirror reflection to the anisotropic case (see~\cite{Brag00}).
According to~\cite{Brag00}, the average $\delta \tilde{u}$ fluctuation size is described as a volume average
of the $\delta u (\vb{r}, t)$ distribution in the characteristic volume of the optical field. The weight each
point of material contributes to the value of $\delta \tilde{u}$ is proportional to the intensity of light in these
points. The optical field amplitude has a gaussian profile in the transverse direction, and presence of
the mirror results in an exponential decay along the axial direction with a characteristic penetration depth
which is the same order of magnitude as the thickness of a typical coating layer.
Therefore, the expression for $\delta u$ is described by
\begin{equation}
 	\delta \tilde{u} = \frac{1}{\pi r_0^2 \ell_p} \int\limits_{-\infty}^{+\infty} \! \dd{x} \, \dd{y} \int\limits_0^{+\infty} \! \dd{z} \; 
    \delta u (\vb{r}, t) \, e^{-(x^2 + y^2) / r_0^2 } e^{-z/\ell_p},
\end{equation}
where $\ell_p$ is the characteristic penetration depth, and $r_0$ is a radius of the incident light beam. 
The resulting spectral density of the volume-averaged temperature is
\begin{equation} \label{eq:Suu_mirror}
 	S_{\tilde{u}\tilde{u}}(\Omega)
 	= \int\limits_{- \infty}^{+\infty} \! \frac{\dd^3 \vb{K}}{(2 \pi)^3} \,
 		\frac{4 \zeta^2 \left( D_{ij} K_i K_j \right) \exp\left[ - \frac{(K_x^2 + K_y^2) r_0^2}{2} \right]}
 		{(1 + K_z^2 \ell_p^2)^2 \left( \Omega^2 + (D_{ij} K_i K_j)^2 \right)}.
\end{equation}
In the thermally isotropic case, it can be shown that it reduces to the results in \citet{Brag00}.
Specifically, if the $1+K_z^2 \ell_p^2$ term is ignored, it reduces to the isotropic result~\cite[\S 3.3.2]{Martin:2013}
\begin{equation}
    S_{\tilde{u}\tilde{u}}(\Omega) \simeq \frac{2 k_\text{B} T^2}{\pi r_0 c_V D} \Re{\int\limits_0^\infty\! \rmd u \frac{u\,\rme^{-u^2 / 2}}{\sqrt{u^2 - \rmi r_0^2 \Omega / D}}}.
\end{equation}

In the thermally anisotropic case, the asymptotic forms of \ref{eq:Suu_mirror} are
\begin{equation}
  S_{\tilde{u}\tilde{u}}(\Omega) = \frac{\zeta^2}{\pi r_0}
  \begin{dcases}
    \frac{\left(2 \Tr \hat{D}_\perp^2 \right)^{-\frac{1}{4}}}{\sqrt{\pi D_{zz}}} \mathop{K}\bigl(\sin \tfrac{\phi}{2} \bigr);
      &  \Omega \ll \frac{D}{r_0^2} \\
    \frac{1}{r_0 \sqrt{2 D_{zz} \Omega}};& \Omega \gg \frac{D}{r_0^2},
  \end{dcases}
\end{equation}
where
\begin{equation}
  \hat D_\perp = 
    \begin{bmatrix}
      D_{xx} & D_{xy} \\
      D_{xy} & D_{yy}
    \end{bmatrix},
\end{equation}
$D$ is the typical diagonal element of this matrix (assumed roughly comparable), $K$ is the complete elliptic integral
of the first kind, and $\cos \phi = (\Tr \hat D_\perp)/\sqrt{2 \Tr \hat D_\perp^2}$.
Note that all these results rely on the beam spot size being larger than the penetration depth (i.e. $r_0 \gg \ell_p$).
Additionally, we note that in the above expressions, the appropriate material parameters to use are those of the substrate;
this amounts to the statement that the temperature fluctuations near the surface of the coating are dominated by the effect
of heat flow in the substrate.

Substituting \cref{eq:Suu_mirror} into \cref{eq:deltaExyrt}
gives us the spectral density of the polarization fluctuations:
\begin{align} 
	\label{eq:tx_final}
	S_{e_x e_x}^\t{t}(\Omega) &= \left|t_x \frac{N}{2}
		\left( \frac{\eps_{xx}^{\prime (1)}}{n_1^2} -\frac{\eps_{xx}^{\prime (2)}}{n_2^2} \right)
		\right|^2 S_{\tilde{u}\tilde{u}}(\Omega),
	\\
	\label{eq:rx_final}
	S_{e_x e_x}^\t{r}(\Omega) &= \left|r_x \frac{\pi}{2}\frac{n_2 \eps_{xx}^{\prime (1)} + n_1 \eps_{xx}^{\prime (2)}}
					{n_1 n_2 (n_1^2 - n_2^2)}
					\right|^2 S_{\tilde{u}\tilde{u}}(\Omega),
	\\
	\label{eq:ty_final}
	S_{e_y e_y}^\t{t}(\Omega) &= \left| t_y \frac{N}{2}
		\left( \frac{\eps_{xy}^{\prime (1)}}{n_1^2} -\frac{\eps_{xy}^{\prime (2)}}{n_2^2} \right)
		\right|^2 S_{\tilde{u}\tilde{u}}(\Omega),
	\\
	\label{eq:ry_final}
	S_{e_y e_y}^\t{r}(\Omega) &= \left|r_y \frac{\pi}{2}\frac{n_2 \eps_{xy}^{\prime (1)} + n_1 \eps_{xy}^{\prime (2)}}
					{n_1 n_2 (n_1^2 - n_2^2)}
					\right|^2 S_{\tilde{u}\tilde{u}}(\Omega).
\end{align}
Note that cross-correlations can be computed the same way and will have the same dependence on
$S_{\tilde{u} \tilde{u} (\Omega)}$. These equations are valid for any crystalline mirror Bragg stack operated near the quarter
wavelength stack condition, for any crystalline material whose in-plane optical anisotropy
is small (i.e $\abs{\Delta n}\ll 1$). They thus describe crystalline mirrors currently being considered
for all precision optical instruments. 

\begin{figure}[t]
	\includegraphics[width=\columnwidth]{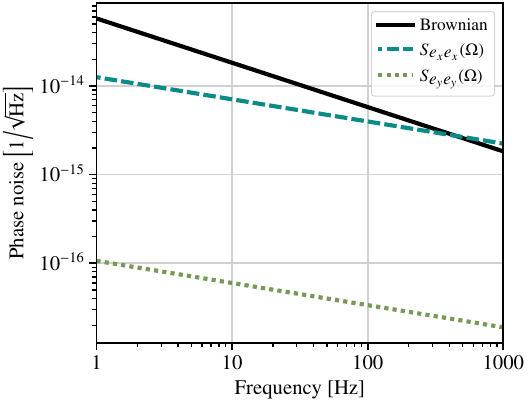}
	\caption{Thermorefringent noise (\cref{eq:rx_final,eq:ry_final}) for an AlGaAs/GaAs high reflector of 50 quarter-wave layers, with \SI{1500}{\nm} light with radius $r_0 = \SI{4}{\cm}$.
    The comparison with the Brownian noise of the coating is also shown.}
	\label{fig:PSD_mirror}
\end{figure}

The plot for the relative power spectral density for one particular coating system (AlGaAs/GaAs) is shown in \cref{fig:PSD_mirror}. The material parameters are given in \cref{tab:material_params}.
Note that the estimate here depicts an alternating stack of identical AlGaAs/GaAs layers, which may not be
optimal from the perspective of reducing thermorefringent noise. As in the case of amorphous coatings, where
the thin-film stack structure can be optimized~\cite{Gor08} to reduce thermo-optic noises, it may also be possible
to optimize the stack structure of crystalline coatings to optimize thermorefringent noise.
The thermorefractive and thermorefringent noises of the coating are also compared to the Brownian noise, which has a phase noise power spectral density
$S_\text{cBr}(\Omega) \simeq (2\pi/\lambda)^2(4k_\text{B} T / \pi r_0^2 \Omega E)(1 - \sigma - 2\sigma^2) \phi d$~\cite{Hong:2012jv,Tait:2020xpg}.
Here the coating thickness is $d = \SI{6}{\um}$ and the coating loss angle is $\phi = \num{1e-5}$.
The approximation symbol indicates that the effect of the light penetration into the coating has been ignored, as has the disparity in the mechanical parameters of the coating and substrate (we chose typical values of $E = \SI{100}{\giga\pascal}$ for the Young modulus and $\sigma = 0.2$ for the Poisson ratio).


\subsection{Thermodynamic beam pointing noise} \label{sec:pointing}

The manifestations of thermodynamic noise considered so far describe the effect of thermal fluctuations
in an anistropic medium on the same spatial mode of the field as the one used to probe the medium. A qualitatively
different effect is that where thermodynamic dielectric fluctuations scatter light from the spatial mode of
the incident field to an orthogonal mode. If the incident field is an transverse mode that is cylindrically
symmetric, and the scattering is predominantly into modes that break that cylindrical symmetry, the effect of scattering
is an apparent change in the angle of the optical beam\,---\,that is, beam pointing noise of thermodynamic origin.

In this section we describe thermodynamic beam pointing noise. A proper accounting of this effect calls for a modal
resolution of the optical field [\cref{eq:dEexpansion}, see also \cref{sec:EMevolution}]:
\begin{equation}
\begin{split}
	\delta E_i(\vb{r},t) = \sqrt{P} e^{\rmi (kn_i z -\omega t)}&\Big\{ f_0(x,y) \delta e_i^{(0)}(z,t) \\ 
		& + \sum_{\alpha\geq 1} f_\alpha (x,y) \delta e_i^{(\alpha)}(z,t) \Big\}.
\end{split}
\end{equation}
where the spatial mode $f_0$ is taken to be the one populated in the incident field, and the higher order modes
$f_{\alpha \geq 1}$ are populated by thermodynamically-induced scattering.
We focus attention on a single higher order mode to which scattering is predominant. 
For example, this captures the common scenario where light in a fundamental Gaussian mode of a laser
($f_0 = \exp [-(x^2+y^2) / (2 r_0^2)] / \sqrt{\pi r_0^2}$) is scattered
into a $(1,0)$ Hermite-Gauss mode ($f_1 =\sqrt{2 / \pi r_0^2} \, x \exp [-(x^2+y^2) / (2 r_0^2)]$).
Since both the modes vary much slower in the transverse direction that along the propagation
direction, arguments similar to the ones in \cref{sec:EMevolution} can be employed to separate out from the
Maxwell equations for the field fluctuations [\cref{eq:dE}], the equations for the polarization components of the relevant higher order
mode. This gives,
\begin{equation}\label{eq:motion_xy_high}
\begin{split}
	\left( \pdv{z} + \frac{n_x}{c}\pdv{t} \right) \delta e_x^{(1)} &=
		\frac{ik \eps_{xx}'}{2n_x}(f_1 | f_0\, \delta u) \langle e_x^{(0)}\rangle \\
	\left( \pdv{z} + \frac{n_y}{c}\pdv{t} \right) \delta e_y^{(1)} &=
		\frac{ik \eps_{xy}' e^{\rmi (n_x -n_y)kz}}{2n_x}(f_1 | f_0\, \delta u) \langle e_x^{(0)}\rangle
\end{split}
\end{equation}
These are very similar to \cref{eq:motion_xy}, except that the stochastic source term on the right-hand side involves
the spatial overlap $(f_1|f_0 \delta u)$ that describes the scattering efficiency from the fundamental mode to the
higher order mode mediated by the temperature field $\delta u$. Employing arguments and techniques similar to the ones
in \cref{sec:anisotropic_transmission}, we calculate the correlation function of the source,
\begin{widetext}
\begin{align}\label{eq:corr_uu_x}
	\avg{ (f_1|f_0 \, \delta u(\vb{r}, t)) (f_1|f_0\, \delta u(\vb{r}', t + \tau)) }
	=  \frac{ \zeta^2 r_0^2 \sum_{\vb{s} \in S} \exp \left[ - \frac{(z - s'_z)^2}{4 D_{zz} |\tau|} \right]}
		{16 \sqrt{\pi^3} \sqrt{D_{zz} |\tau|} 
		\sqrt{ \left( 2 D_{xx} |\tau| + r_0^2 \right)^3 \left( 2 D_{yy} |\tau| + r_0^2 \right) }}.
\end{align}
\end{widetext}
This fixes the statistical properties of the source that drives \cref{eq:motion_xy_high}. Since the latter is
structurally similar to the equations of motion that describe the transmission problem in \cref{sec:anisotropic_transmission},
they can be solved similarly. We thus arrive at the spectral density of the polarization fluctuations in the higher order
mode,
\begin{align}
	S_{e_x^{(1)} e_x^{(1)}} &= \frac{\zeta^2 k^2 |\eps_{xx}'|^2  \ell r_0^2}{64 \pi n_x^2 \sqrt{D_{yy} D_{xx}^3}} 
		I^{(1)}(\Omega, n_x \, \Omega)
	\\
	S_{e_y^{(1)} e_y^{(1)}} &= \frac{\zeta^2 k^2 |\eps_{xy}'|^2 \ell r_0^2}{64 \pi n_y^2 \sqrt{D_{yy} D_{xx}^3}} 
		I^{(1)}(\Omega, n_x \omega + n_y \Omega),
\end{align}
where,
\begin{align}\label{eq:Integral_x}
	I^{(1)}(\Omega_1, \Omega_2) = \int\limits_0^{\infty}\! \dd\tau \, \frac{\cos \Omega_1 \tau \; 
    \exp \left[ - \frac{\Omega_2^2}{c^2} D_{zz} \tau \right]}{ \sqrt{( \tau + \tau_x)^3 (\tau + \tau_y)}}.
\end{align}
is analogous to the integral in \cref{eq:Iintegral}.
For Fourier frequencies that are small compared to the
thermal diffusion timescale (i.e. $D_z \tau_+ /\ell^2  \ll \Omega \tau_+ \ll 1$), 
the required limiting expressions for $I^{(1)}$ are given by,
\begin{align*} \label{eq:final_int_x}
	I^{(1)}(0, 0) = \frac{2}{\sqrt{\tau_x} \left( \sqrt{\tau_x} + \sqrt{\tau_y} \right)}
	\\
	I^{(1)}(0, n_x k c) = \frac{1}{ k^2 n_x^2 D_{zz} \sqrt{\tau_x^3 \tau_y}}.
\end{align*}
Using these, we have the power spectral density of the polarization fluctuations of the higher order mode,
\begin{equation}\label{eq:Sxy_scattering}
\begin{split}
	S_{e_x^{(1)} e_x^{(1)}} &= \frac{\zeta^2 k^2 |\eps_{xx}'|^2 \ell}{32 \pi n_x^2 \sqrt{D_{xx}}
			\left( \sqrt{D_{xx}} + \sqrt{D_{yy}} \right)},\\
	S_{e_y^{(1)} e_y^{(1)}} &= \frac{\zeta^2 k^2 |\eps_{xy}'|^2  \ell}{32 \pi n_x^2 n_y^2 D_{zz} r_0^2},
\end{split}
\end{equation}
which are both white noise at these low frequencies.

\subsubsection{Thermodynamic pointing noise in amorphous media}

The above equations predict that even for an amorphous medium, beam-pointing noise
due to thermodynamically-mediated scattering into higher order modes can exist.
Indeed, in general~\cite{And84}, scattering of light from the (0,0) Gaussian mode into 
the (1,0) or (0,1) Hermite-Gauss mode is equivalent to beam pointing noise by an angle
$\delta \psi = (k r_0)^{-1} (f_1|\delta E_x)/\sqrt{P}$. Thus the spectral density of the
beam pointing angle fluctuations is given by $S_{\psi \psi} = (kr_0)^{-2} S_{e_x^{(1)}e_x^{(1)}}$.
For an amorphous medium, characterized by an isotropic thermal conductivity $D_{ij}=\delta_{ij}D$
and an isotropic dielectric constant
\begin{equation*}
	\eps_{ij} = \delta_{ij} \left( n + \pdv{n}{T}  \right)^2 
	\approx \delta_{ij} \left( n^2 + 2n \pdv{n}{T} \right),
\end{equation*}
where $n$ is the (average) refractive index, \cref{eq:Sxy_scattering} reduces to
\begin{equation}
	S_{e_x^{(1)} e_x^{(1)}}(\Omega) = \frac{(\zeta k)^2 \ell}{16 \pi D}\left( \pdv{n}{T} \right)^2 \!.
\end{equation} 
Referring these to beam pointing angle, we find
\begin{equation}
	S_{\psi \psi}(\Omega) = \frac{\zeta^2 \ell}{16 \pi r_0^2 D}\left( \pdv{n}{T} \right)^2 \!.
\end{equation}
For the geometry considered previously ($r_0 = \SI{100}{\um}$, $\ell = \SI{1}{\cm}$), the
pointing fluctuation in cryogenic silicon (\cref{tab:material_params}) is of order \SI{e-13}{\radian/\sqrt{\Hz}};
for room-temperature fused silica,\footnote{For silica, we take $\partial n/\partial T = \SI{11}{ppm\ \kelvin^{-1}}$, $\kappa = \SI{1.4}{\watt\ \meter^{-1}\ \kelvin^{-1}}$, and $c_V = \SI{1.6e6}{\joule\ \meter^{-3}\ \kelvin^{-1}}$.} it is of order \SI{e-12}{\radian/\sqrt{\Hz}}.


\section{Manifestation of polarization noise in optical detection}\label{sec:detection}

The previous sections establish the formalism, and then use it to determine polarization fluctuations
in optical fields due to their interaction with crystalline optical materials in thermal equilibrium. 
The precise manner in which these polarization fluctuations manifest in signals that are typically measured in an 
experiment is the concern of this section.

\subsection{Direct photodetection}

We will consider, as before, that the electric field in the plane transverse to
the propagation direction is of the form [\cref{eq:avgE,eq:dEexpansion}, with mode indices dropped
under the assumption that we limit attention to a single spatial mode; i.e. neglecting beam pointing noise]
\begin{equation}\label{eq:det:Ei}
\begin{split}
    E_i(\vb{r},t) &= \avg{E_i(\vb{r},t)} + \delta E_i(\vb{r},t), \qquad (i=x,y)
\end{split}
\end{equation}
where,
\begin{equation}\label{eq:det:EdE}
\begin{split}
    \avg{E_i} &= \sqrt{P} e^{\rmi (k n_i z -\omega t)} f(x,y) \langle e_i\rangle\\
    \delta E_i &= \sqrt{P} e^{\rmi (k n_i z -\omega t)} f(x,y)\delta e_i(z,t).
\end{split}
\end{equation}
This field is incident on a photo-emissive surface, held perpendicular to the
propagation direction, at $z=\ell$. The photocurrent emitted by the detector is then~\cite{MandSud64}
\begin{equation*}
    I(t) = \alpha \int\limits_\mathcal{D} \dd^2\vb{r} \, E_i^*(\vb{r},t)E_i(\vb{r},t) \,,
\end{equation*}
where $\alpha$ is the responsivity, $\mathcal{D}$ is the domain of the photoemissive surface, 
$\vb{r}=(x,y,\ell)$, and $\dd^2\vb{r} = \dd{x}\,\dd{y}$. 
We will assume that the
the area of $\mathcal{D}$ is much larger than the transverse extent of the electric fields
involved so that the optical beam is not clipped; we will thus extend
the above integrals to the entire $xy$ plane. Using \cref{eq:det:Ei} in the above
equation, and neglecting terms second order in the electric field fluctuations,
the photocurrent splits into a mean (``DC'') part and a fluctuating (``AC'') component:
\begin{equation}
\begin{split}
    I(t) &\approx \avg{I} + \delta I(t)\\
    \avg{I} &= \alpha \int \dd^2 \vb{r} \, \abs{\avg{E_i}}^2 \\
    \delta I &= \alpha \int \dd^2 \vb{r} \, \left( \avg{E_i^*}\, \delta E_i + \t{c.c.} \right) \,.
\end{split}
\end{equation}
Since $\delta I$ is linear in $\delta E_i$, and the latter is a Gaussian stochastic process, so is the former. 
Thus the statistical properties of the photocurrent are fully characterized by the two-time correlation
function, $C_{II}(t,\tau) = \avg{\delta I(t) \delta I(t+\tau)}$.
Using the explicit form of $\delta I$, its correlation function can be written as
\begin{widetext}
\begin{multline}
    C_{II}(t,\tau) = \alpha^2 \int \dd^2 \vb{r} \, \dd^2 \vb{r}'
    \left\{\left[
        \avg{E_i^*(\vb{r},t)} \avg{\delta E_i(\vb{r},t)\delta E_j^* (\vb{r}',t+\tau)} \avg{E_j (\vb{r}',t)} + \t{c.c.} 
    \right] \right.\\
    \left. + \left[ 
        \avg{E_i^*(\vb{r},t)} \avg{\delta E_i(\vb{r},t)\delta E_j (\vb{r}',t+\tau)} \avg{E_j^* (\vb{r}',t)} + \t{c.c.} 
     \right] \right\}.
\end{multline}
All four terms here are independent of the optical frequency $\omega$, so that the statistical properties of
the photocurrent are independent of the optical carrier. The first two terms are further only sensitive to
stationary fluctuations of the field, whereas the second pair are sensitive to non-stationary fluctuations as well.
We neglect the second pair of terms since field fluctuations due to thermorefringent noise are stationary. 
Then the correlation function only depends on the time delay $\tau$; so we use the notation, 
$C_{II}(\tau) = C_{II}(t,\tau)$. 
Introducing the two-point correlation function of the electric field,
\begin{equation}
\begin{split}
    C_{E_i E_j}(\vb{r}',\tau) &\equiv \avg{\delta E_i (\vb{r},t) \delta E_j^* (\vb{r}+\vb{r}',t+\tau)} \\
\end{split}
\end{equation}
we have
\begin{equation}\label{eq:det:CIIstat}
    C_{II}(\tau) = \alpha^2 \int \dd^2 \vb{r} \, \dd^2 \vb{r}' \left[ 
        \avg{E_i^*(\vb{r},t)} C_{E_i E_j}(\vb{r}-\vb{r}',\tau) \avg{E_j (\vb{r}',t+\tau)} + \t{c.c.}
    \right].
\end{equation}
\end{widetext}

Using the explicit form of the field fluctuations in \cref{eq:det:EdE}, we have that,
\begin{equation*}
    C_{E_i E_j}(\vb{r}',\tau) = P\, f(\vb{r})f^*(\vb{r}+\vb{r}')\, C_{e_i e_j}(\tau),
\end{equation*}
where $C_{e_i e_j}(\tau) \equiv \avg{\delta e_i(\ell,t) \delta e_j^* (\ell,t+\tau)}$, is the correlation
function of the vectorial polarization fluctuations. 
Inserting the expression for the mean field from \cref{eq:det:EdE} in \cref{eq:det:CIIstat}, 
the spatial integral in \cref{eq:det:CIIstat} factorizes out, which gives a numerical constant
that can be absorbed by redefining the responsivity $\alpha$ (and in fact describes the geometric contribution to the
detection efficiency); we thus arrive at,
\begin{equation}
     C_{II}(\tau) = \alpha^2 P^2 \, \left[ \langle e_i^*\rangle C_{e_i e_j}(\tau) \langle e_j\rangle + \t{c.c.} \right].
 \end{equation} 
Finally, stationary photocurrent fluctuations can be equivalently described by 
the Fourier transform of its two-time correlation function, the power spectral density, 
$S_{II}(\Omega) = \int C_{II}(\tau) e^{i \Omega \tau}\, d \tau$, which assumes the form,
\begin{equation}\label{eq:det:SiPD}
\begin{split}
    S_{II}(\Omega) 
    = (\alpha P)^2 \left[ \langle e_i^*\rangle S_{e_i e_j} (\Omega) \langle e_j\rangle + \t{c.c.}\right].
\end{split}
\end{equation}

These photocurrent fluctuations can be referred to relative intensity noise of the optical field, 
$S_{II}/(\alpha P)^2 = \langle e_i^*\rangle S_{e_i e_j} (\Omega) \langle e_j\rangle + \t{c.c.}$. Thus, 
when a depolarized field is subjected to direct photodetection, thermorefringent noise manifests as 
apparent intensity noise. That is one operational interpretation of the polarization noise plotted
in \cref{fig:PSD,fig:PSD_mirror}.

Note that the photocurrent fluctuations emitted by subjecting a depolarized beam to direct photodetection
does not allow inference of the full polarization covariance matrix $C_{\vb{ee}}$ (and therefore its
Fourier transform $S_{\vb{ee}}$).
In particular, for a choice of the input carrier polarization $\langle \vb{e} \rangle$, the photocurrent spectrum [\cref{eq:det:SiPD}] 
is a linear combination of the elements of $S_{\vb{ee}}$, from which the full matrix cannot be reconstructed. 
Indeed, attempts to assemble a set of measurements, by varying the mean input polarization, that is linearly
independent in the elements of $S_{\vb{ee}}$ is not guaranteed to succeed in general, since changing the input polarization
can change the transduction of the noise properties of the sample being interrogated (see \cref{fig:PSD}, for example).

\subsection{Balanced homodyne polarimetry}\label{sec:det:homodyne}

The most general type of optical detection that a polarized state of the optical field can be subjected to is
balanced homodyne polarimetry. Here, the signal\,---\,the depolarized output of a system, represented by the electric
field $E_i$ in \cref{eq:det:Ei}\,---\,is mixed with a local oscillator (LO) in a pure and controllable polarization state 
that has a well-defined and controllable phase difference with the signal at a balanced polarizing beam-splitter; 
the resulting outputs are photodetected and their photocurrrent subtracted. 
We will show that by controlling the local oscillator polarization and phase, the subtracted photocurrent can be used 
to deduce the spectral covariance matrix $S_{\vb{ee}}$ of the signal without changing the optical field used to probe the system.

We assume that the LO is prepared in the same transverse spatial mode $f$, and longitudinal
mode with wave-vector $k$, as the signal of interest, so we take its electric field to be given by,
\begin{equation}\label{eq:det:LOfield}
	E_i' = \avg{E_i'} = \sqrt{P'} e^{\rmi (k_i z - \omega t)} f(x,y) \langle e_i'\rangle,
\end{equation}
where $P' \gg P$ is the local oscillator power and $e_i'$ its mean polarization.
The assumption that the LO power is much larger than that of the signal effectively means that polarization 
fluctuations in the LO can be neglected, which is tacit in the above ansatz and in all that follows.
This field is superposed with the signal at a balanced beam-splitter; the fields at its output are given by,
\begin{equation}
\begin{split}
	E^\pm_i = \frac{1}{\sqrt{2}} (E_i' \pm E_i)
		\approx \frac{1}{\sqrt{2}} \left(\avg{E_i'} \pm \delta E_i \right),
\end{split}
\end{equation}
where the second equality uses the fact that the LO is overwhelmingly more powerful than the signal (i.e. $P' \gg P$) and
so neglects a term of order $\sqrt{P/P'}$.
Each of the outputs is passed through a polarization analyzer (``polarizer'') which projects the polarization vector
onto a chosen direction; this can be modelled by the transformation,
\begin{equation}
	E_i^\pm \mapsto J_{ij}^\pm E_j^\pm,
\end{equation}
where the projective nature of the polarizer implies that the Jones matrices $\vb{J}^\pm$ satisfy
$\vb{J}^\pm =(\vb{J}^\pm)^\dagger = (\vb{J}^\pm)^2$.
These fields are individually detected, producing the photocurrents,
\begin{equation}
	I^\pm = \alpha \int\limits_\mathcal{D} \dd^2\vb{r} \, (J_{ij}^\pm E_j^\pm)^* (J_{ik}^\pm E_k^\pm),
\end{equation}
where the integrands are evaluated at the detector plane $z=\ell$.
Combining the above equations it can be shown that the fluctuations in these photocurrents are given by,
\begin{equation*}
	\delta I^\pm = \pm \frac{\alpha}{2}\int\limits_\mathcal{D} \dd^2\vb{r} \left[ \avg{E_i'}^* J^\pm_{ij}\, \delta E_j + \t{c.c.} \right].
\end{equation*}
The individual photocurrents are subtracted to produce the homodyne photocurrent $I = I^+ - I^-$, whose
fluctuations assume the form,
\begin{equation}
	\delta I = \frac{\alpha}{2}\int\limits_\mathcal{D} \dd^2\vb{r} \left[ \avg{E_i'}^* (J_{ij}^+ + J_{ij}^-)\, \delta E_j + \t{c.c.} \right].
\end{equation}
In order to maximize the sensitivity of the subtracted photocurrent to fluctuations in the signal field, it 
is best to choose polarizers that are orthogonal, in which case $\vb{J}^+ \vb{J}^- = \vb{0}$ 
and $\vb{J}^+ + \vb{J}^- = \vb{1}$. Physically, this choice corresponds to the intuition that each photodetector
be sensitive to polarization fluctuations in orthogonal directions, so that their equal-weight superposition
contains full information of both polarization components \footnote{In fact, complete information is contained
in any linearly independent combination of polarization states, but in this case, the photocurrents will need to be
combined with unequal weights.}. With this choice, the homodyne photocurrent simplifies to,
\begin{equation}\label{eq:det:dIhom}
	\delta I = \frac{\alpha}{2}\int\limits_\mathcal{D} \dd^2\vb{r} \left[ \avg{E_i'}^* \delta E_i + \t{c.c.} \right],
\end{equation}
similar to the case of direct photodetection, except that the signal field fluctuations that are 
transduced are the ones that lie along the polarization of the mean LO field.

Inserting the explicit forms of the LO and signal fields [\cref{eq:det:EdE,eq:det:LOfield}],
the homodyne photocurrent fluctuations in \cref{eq:det:dIhom} becomes,
\begin{equation}\label{eq:det:dIhom_de}
	\delta I = \frac{\alpha}{2}\sqrt{P P'} \left[ \langle e_i'\rangle^* J_{ij} \delta e_j + \t{c.c.} \right],
\end{equation}
where $\phi$ is the common difference between the longitudinal modes of the LO and signal, and 
\begin{equation*}
	\vb{J} = \begin{bmatrix} e^{\rmi k(n_x -1)\ell} & 0 \\ 0 & e^{\rmi k(n_y -1)\ell} \end{bmatrix}
\end{equation*}
is the Jones matrix describing the phase retardation between the LO and signal carrier polarizations as they
propagate through to the photodetectors. Indeed by setting $\langle e_j''\rangle = J_{ji}^* \langle e_i'\rangle $, the photocurrent
fluctuations can be seen to be proportional to $\langle e_j''\rangle^* \delta e_j + \t{c.c.}$, where 
$\langle e_j''\rangle$ can be identified
with the polarization state of the LO after passing through a phase retarder described by the Jones matrix $\vb{J}^\dagger$.
In this sense, if the LO polarization state is 
completely controllable, the effect of $\vb{J}$ can in principle be absorbed into the definition of $\vb{e}'$;
we do so in the following.
Computing the two-time correlation of the photocurrent fluctuations in \cref{eq:det:dIhom_de}, omitting terms that
are non-stationary, and computing the Fourier transform, gives the photocurrent spectral density,
\begin{equation}\label{eq:det:SiHom}
 	S_{II}(\Omega) = \left( \frac{\alpha}{2}(PP')^{1/2} \right)^2 \left[ 
 	\langle e_i'\rangle^* S_{e_i e_j}(\Omega) \langle e_j'\rangle + \t{c.c} \right].
\end{equation} 
In contrast with the case of direct photodetection [\cref{eq:det:SiPD}], by changing the LO polarization $\vb{e}'$, all
elements of the spectral covariance matrix of the signal polarization can be measured without perturbing the field
incident on the sample.

Note that in general polarization fluctuations contaminate the homodyne photocurrent in all quadratures. 
To see this, re-introduce the phase retardation between the LO and signal, $\avg{e_i'} \rightarrow 
\avg{e_i'} e^{\rmi \phi_i}$, and notice that whatever value of the relative phase $\phi_i$ is chosen, the photocurrent
spectral density $S_{II}$ is generically non-zero. In this sense, thermorefringent noise can limit the sensitivity
of an interferometric measurement in all quadratures. This is nothing but the manifestation of the fact that the
noisy polarization state of the signal cannot perfectly interfere with the pure-polarized LO --- a fact that is independent
of signal quadrature. 

\subsection{Coherent cancellation of thermorefringent noise in signal detection}\label{sec:cohCan}

In the context of sensitive polarimetry experiments, the fact that thermorefringent noise is always
lesser in the polarization state orthogonal to the probe field, i.e. $S_{e_y e_y} < S_{e_x e_x}$, suggests
arranging the experiment so that the signal of interest is produced in that polarization, 
i.e. $\delta e_y^\t{sig}$. The resulting signal from a balanced homodyne polarimeter with LO state
$\vb{e}' = (\cos \theta, e^{i \phi} \sin \theta)$ is
\begin{equation}\label{eq:cc:Si}
\begin{split}
    S_{II}^\t{hom} &\propto S_{e_y e_y}^\t{sig} +\Big[ S_{e_y e_y} + \cot^2(\theta) S_{e_x e_x}\\
        &\hphantom{=} \hspace{5em} + 2\cot(\theta) \Re{e^{i \phi} S_{e_x e_y}^*}\Big]\\
        & \equiv S_{e_y e_y}^\t{sig} + S_{e_y e_y}^\t{app},
\end{split} 
\end{equation} 
which is \cref{eq:det:SiHom} referred to the polarization signal of interest.
The terms in the brackets in the first line represent the apparent signal arising from 
thermorefringent noise, denoted $S_{e_y e_y}^\t{app}$.
The primary objective of any polarimetry experiment is the maximization of the signal-to-noise 
ratio $S_{e_y e_y}^\t{sig}/S_{e_y e_y}^\t{app}$; equivalently, the minimization of the noise
$S_{e_y e_y}^\t{app}$ once the signal is fixed.

If there existed no correlations between thermorefringent noise of orthogonal polarizations 
(i.e. $S_{e_x e_y} =0$), then, $S_{e_y e_y}^\t{app} = S_{e_y e_y} + \cot^2(\theta) S_{e_x e_x}$.
This can be minimized by choosing LO tuned to the signal polarization, i.e. $\theta = \pi/2$, in which 
case the sensitivity to signal polarization is limited by thermorefringent noise in the same polarization
(i.e. $S_{e_y e_y}$).
Indeed this signal extraction strategy is conventionally practised for a different reason: to avoid extraneous 
background from the probe field. 

However, since thermorefringent noise is correlated across the probe and signal polarizations 
(i.e. $S_{e_x e_y} \neq 0$, as seen in \cref{fig:PSD}), 
better signal extraction strategies that harness these correlations can be imagined. 
Mathematically, the LO polarization angles $(\theta,\phi)$ can be chosen so that the negative values of the 
correlation terms in $S_{e_y e_y}^\t{app}$ cancel with the positive terms.
Expressing $S_{e_y e_y}^\t{app}$ by completing squares on $\cot{\theta}$, we find
\begin{equation}\label{eq:cc:Sapp}
\begin{split}
    S_{e_y e_y}^\t{app} &= 
        S_{e_y e_y} - \frac{\Re{e^{i \phi} S_{e_x e_y}^*}^2}{S_{e_e e_x}}\\
    &\qquad + \left[\cot \theta \cdot S_{e_x e_x} + \Re{e^{i \phi} S_{e_x e_y}^*} \right]^2.
\end{split}
\end{equation}
It is clear that this is minimized at a Fourier frequency of interest $\Omega$ when the second
term is maximized by proper choice of $\phi$, and the third term is nulled by choice of $\theta$.
Noting that $\Re{e^{i \phi} S_{e_x e_y}^*}^2 = \abs{S_{e_x e_y}^2} \cos^2(\phi -\arg S_{e_x e_y})$,
these optimal choices are
\begin{equation}\label{eq:cc:opt}
\begin{split}
    \phi_\t{opt} (\Omega) &= \arg S_{e_x e_y}[\Omega]\\
    \theta_\t{opt}(\Omega) &= \cot^{-1} \frac{-\Re{e^{i \phi} S_{e_x e_y}^*[\Omega]}}{S_{e_x e_x}[\Omega]}.
\end{split}
\end{equation}
With this choice, the noise at that frequency is
\begin{equation}
    S_{e_y e_y}^\t{app}(\Omega)\vert_{\theta_\t{opt},\phi_\t{opt}} = 
    S_{e_y e_y} (\Omega) \left[ 1- \frac{\abs{S_{e_x e_y}^2 (\Omega)}}{S_{e_x e_x}(\Omega) S_{e_y e_y}(\Omega)} \right].
\end{equation}
Since the correlation is bounded by the Cauchy--Schwarz inequality 
$|S_{e_x e_y}^2| \leq S_{e_x e_x} S_{e_y e_y}$, in principle, perfect cancellation at a desired
Fourier frequency is possible if the correlations are perfect (i.e. saturate the inequality).
Even with imperfect correlations, narrow-band evasion of thermo-refringent noise is possible via
balanced homodyne polarimetry via coherent cancellation. This strategy always outperforms --- in a
narrow-band of choice --- the conventional signal extraction strategy of tuning the LO to a polarization
orthogonal to the probe.

\section{Conclusions}

Having emerged from the thicket, we can now contextualize thermorefringent noise in the wider landscape
of thermo-optic noises.
Fluctuations of the apparent temperature of amorphous materials cause their optical properties to
fluctuate, which can manifest as extraneous noise in precision optical measurements~\cite{Harry12}.
The most insidious source of such thermo-optic noise is that due to fluctuations in the thickness
of coatings on mirrors, the intensity of which is related to the mechanical loss of these materials.
Driven by the idea that it is the glassy energy landscape of amorphous materials that gives rise
to mechanical dissipation~\cite{Phil87,Pohl02,BasRow13,TrinChen16}, a concerted effort to discover more pristine materials 
has ensued in communities engaged in precision optical measurements. Recent measurements~\cite{Cole13,ChalAdhi16}
have unearthed evidence that crystalline materials may offer some refuge from thermo-optic noises
because of the absence of glassy behavior. In the current study, we have demonstrated that precisely
because of the anisotropy of the crystalline state, qualitatively novel sources of thermodynamically
driven optical noises can arise. 

In particular, fluctuations in temperature can be anisotropic, which drive fluctuations in the
dielectric tensor of the medium, resulting in the polarization of an incident optical field
to transmute into an impure state. We term this thermorefringent noise.
An impure polarization state manifests in optical measurements
via its inability to interfere perfectly with a reference pure-polarized field. 
The result is that thermorefringent noise can manifest as apparent noise in any quadrature of the optical field,
quite unlike thermo-optic noise from amorphous media.
There are also other manifestations of
thermorefringent noise, such as the thermal scattering of light into orthogonal polarizations, which can
be detrimental to precision polarimetry experiments. 
In addition, we also discover that thermodynamic
scattering into higher-order spatial modes is possible, even in amorphous optical media.

The phenomenology of thermorefringent noise critically depends on the temperature-dependent parts
of the dielectric tensor, which can in turn depend on residual stresses on optical materials such as coatings.
These poorly understood aspects of such materials need to be carefully characterized to ascertain
the realistic limits that thermorefringent noise will place on precision optical measurements.

We have also proposed a novel signal extraction strategy
employing balanced homodyne polarimetry which can coherently cancel thermo-refringent noise. This
technique crucially relies on the complete theoretical understanding of thermo-optic noises that our
formalism has captured, including thermodynamically induced correlations in the optical polarization.
The coherent cancellation strategy can evade correlated polarization noise in a narrow frequency of
choice by simple tuning of the local oscillator polarization state, and is only limited by the 
strength of the correlations.

\section{Acknowledgements}

We thank Sergey Vyatchanin for a discussion about the fluctuation-dissipation
theorem for the heat equation, and Matt Evans for motivating us to reconcile our approach with
that via Levin's ``direct approach'' (the result is \cref{sec:levin}).
EDH is supported by the MathWorks, Inc. This work has document number
LIGO--P2100419.

\bibliography{refs_TReN}


\appendix

\section{Directional correlation of thermal noise}\label{sec:norm_constant}

In this appendix we consider the general problem of reconciling the microscopic anisotropic description
of the temperature field $u$, given in \cref{eq:Tsystem,eq:corr_q}, with the macroscopic thermodynamic expectation
for the temperature, given in \cref{eq:VarT}.

If we assume that the noise $\zeta_i(\vb{r},t)$ is uncorrelated across space and time, the only remaining
source of correlation are directional. Since there are only three second rank tensors dictated by the system, 
namely $\delta_{ij}, D_{ij}, (D^{-1})_{ij}$, any directional correlation must be captured in the general expression
\begin{multline}\label{eq:general_xi_noise}
	\avg{\zeta_i (\vb{r},t) \zeta_j (\vb{r}',t')} = 
	(\zeta_0^2\, \delta_{ij} + \zeta_1^2 \, D_{ij} + \zeta_{-1}^2 \, D_{ij}^{-1})	\\			
	\times \delta(\vb{r}-\vb{r}')\, \delta(t-t'),
\end{multline}
where $\zeta_{-1,0,1}$ are scalars to be determined.

According to \cref{eq:diffusionT} the local temperature $u$ is driven by the noise $\eta = \partial_i \zeta_i$. 
The above choice for $\zeta_i$ implies that the Fourier transform, \linebreak $\eta(\vb{K},\Omega) 
= \int \dd{\vb{r}}\, \dd{t} \, \eta(\vb{r},t) e^{-i(\vb{K}\cdot \vb{r}- \Omega t)}$, is characterized by
\begin{multline}
	\avg{\eta (\vb{K}, \Omega) \eta^* (\vb{K}', \Omega')} = \\
		=  (\zeta_0^2\, K_i K_j + \zeta_1^2 \, D_{ij} K_i K_j + \zeta_{-1}^2 \, D^{-1}_{ij}K_i K_j) \, \times
 		\\ \times (2 \pi)^4 \, \delta(\vb{K}-\vb{K}')\, \delta(\Omega-\Omega').
\end{multline}
Since the relation between the local temperature $u$ and $\eta$ is linear (\cref{eq:diffusionT}), it can be
solved via a Fourier transform to produce,
\begin{multline}
	\avg{u (\vb{K}, \Omega) u^* (\vb{K}', \Omega')} = \\
		\frac{\avg{\eta(\vb{K}, \Omega) \eta^*(\vb{K}', \Omega')}}{(-i\Omega + D_{ij}K_iK_j)(i\Omega' + D_{ij}K'_i K'_j)}.
\end{multline}

The scalars $\zeta_{-1,0,1}$ that determine the nature of the directional correlation of the noise $\zeta_i$
(\cref{eq:general_xi_noise}) need to be such that the thermodynamic relation for the macroscopic temperature
(\cref{eq:VarT}), $\Var{T} = k_B T^2/C_V$ is consistent with the volume-average of the microscopic temperature
$u$. That is, we demand,
\begin{multline} \label{eq:int_for_coef}
	  \frac{k_B T^2}{C_V} = \Var{T} =  \\ = \frac{1}{V} \int\limits_V \dd{V} \; \frac{1}{V'} \int\limits_{V'} \dd{V'} \; \avg{u (\vb{r}, t) u^* (\vb{r}', t)}.
\end{multline}
where
\begin{multline} \label{eq:in_to_besolved}
	 \avg{u (\vb{r}, t) u(\vb{r}', t)} = \int \frac{\dd\vb{K} \, \dd\Omega \, \dd\vb{K}' \, \dd\Omega'}{(2 \pi)^8} \times \\
	\avg{u (\vb{K}, \Omega) u^* (\vb{K}', \Omega')} e^{i(\Omega - \Omega')t}
    e^{i(\vb{K}\cdot \vb{r} - \vb{K}'\cdot \vb{r}')},
\end{multline}
The integral in equation \cref{eq:in_to_besolved} can be reduced to,
\begin{multline} \label{eq:int_norm}
	 \avg{u (\vb{r}, t) u(\vb{r}', t)} = \int \frac{\dd\vb{K}}{(2 \pi)^3} 
     \frac{e^{i\vb{K}\cdot (\vb{r} - \vb{r}')}}{2}\times \\ 
	 \frac{(\zeta_0^2\, K_i K_i + \zeta_1^2 \, D_{ij} K_i K_j 
    + \zeta_{-1}^2 \, D_{ij}^{-1}K_j K_i )}{D_{ij} K_i K_j} .
\end{multline}
The fraction in the second line of the integrand has an essential discontinuity at 
$\vb{K} = 0$, unless $\zeta_0 = \zeta_{-1} = 0$ or $D_{ij}$ is proportional to identity matrix. 

The discontinuity dictates the value of the integral, and we will show that the integral
is multivalued, unless there is no discontinuity. In case of isotropic medium, 
the integral in \cref{eq:int_norm} returns a value proportional to $\delta(\vb{r} - \vb{r}')$ 
(see Appendix B in~\cite{Brag99}). 
In that case, asymptotics of the integrand at $\vb{K} = 0$ defines the proportionality coefficient $\zeta_1^2$. 
In the case of anisotropic medium, the scalars $\zeta_{0,1}$ can in principle contribute. 
To study their contributions, we will perform the integral. First, we separate the integral into three 
terms:
\begin{multline} \label{eq:int_form}
	 \avg{u (\vb{r}, t) u(\vb{r}', t)}_\alpha = \int \frac{\dd\vb{K}}{(2 \pi)^3} \, \frac{F_\alpha(\vb{K})}{2} \, e^{i\vb{K}\vb{r} - i\vb{K}'\vb{r}'},
\end{multline}
where, $F_0(\vb{K}) = \zeta_0^2 \, K_i K_i / D_{ij} K_i K_j$, the second integrand
$F_1(\vb{K}) = \zeta_1^2 \, D_{ij} K_i K_i / D_{ij} K_i K_j = \zeta_1^2$, and the third,
$F_{-1}(\vb{K}) = \zeta_{-1}^2 \, D^{-1}_{ij} K_i K_i / D_{ij} K_i K_j$.
Then the correlation in the physical temperature fluctuations can be expressed as the sum,
\begin{equation}
     \avg{u (\vb{r}, t) u(\vb{r}', t)} = \sum_{\alpha\in \{0,\pm 1\}} \avg{u (\vb{r}, t) u(\vb{r}', t)}_\alpha.
\end{equation} 
We now compute each of the terms in the sum.

The integral for $F_1(\vb{K})$ is reduced to a $\delta$-function, since $F_1(\vb{K})$ doesn't have an essential discontinuity at $\vb{K} = 0$:
\begin{equation}
	\avg{u (\vb{r}, t) u(\vb{r}', t)}_1 = \frac{\zeta_1^2}{2} \, \delta(\vb{r} - \vb{r}').
\end{equation}

The integrals containing $F_{0,1}(\vb{K})$ are more complicated, but both of them are computable in the same 
fashion, namely by changing variables to generalized spherical coordinates. For easier pedagogy we perform the variable substitution in steps. 
As a first step, we consider the basis in which $D_{ij}$ is diagonalized (it can be, since it is symmetric)
with eigenvalues $D_i$.
We perform a substitution $\tilde{K}_i = \sqrt{D_i} K_i$, $v_i = (r_i - r'_i) / \sqrt{D_i}$ to get rid 
of $D_{ij}$ in the denominator. 
Then we rotate the resulting coordinate system so as to make $\vb{v}$ lie along the z-direction of the rotated
system. Finally, the $\tilde{\vb{K}}$ integration is performed in spherical coordinates. 
The integral containing $F_0$ then takes the form
\begin{widetext}
\begin{equation*} 
	 \avg{u (\vb{r}, t) u(\vb{r}', t)}_0 = 
	\frac{\zeta_0^2}{2} \int \frac{\tilde{K}^2 \sin{\theta}\, \dd\tilde{K}\, \dd\theta}{(2 \pi)^2 \sqrt{\det \hat{D} }} e^{i\tilde{K}v\cos \theta} \\ 
	\times \left[ \frac{1}{2}\left(\frac{1}{D_x} + \frac{1}{D_y}\right) \sin^2 \theta + \frac{1}{D_z} \cos^2 \theta \right].
\end{equation*}
Performing the polar integral gives
\begin{equation} \label{eq:int0_form}
	 \avg{u (\vb{r}, t) u(\vb{r}', t)}_0 = \frac{\zeta_0^2}{2} \left\{ \left( 4 D_z^{-1} - \text{Tr}(\hat{D}^{-1}) \right) \delta(\vb{r} - \vb{r}') 
		- \frac{3 D_z^{-1} - \text{Tr}(\hat{D}^{-1})}{4 \pi \sqrt{\text{det}\hat{D}}} \times \frac{1}{\left( D^{-1}_{ij} (r_i - r'_i)(r_j - r'_j) \right)^{3/2}}\right\}.
\end{equation}
The integral containing $F_{-1}(\vb{K})$ can be computed applying the same method: 
\begin{equation} \label{eq:int-1_form}
     \avg{u (\vb{r}, t) u(\vb{r}', t)}_{-1} = \frac{\zeta_{-1}^2}{2} \left\{ \left( 4 D_z^{-2} - \text{Tr}(\hat{D}^{-2}) \right) \delta(\vb{r} - \vb{r}')
        - \frac{3 D_z^{-2} - \text{Tr}(\hat{D}^{-2})}{4 \pi \sqrt{\text{det}\hat{D}}} \times \frac{1}{\left( D^{-1}_{ij} (r_i - r'_i)(r_j - r'_j) \right)^{3/2}} \right\}.
\end{equation}
\end{widetext}

Note however that the expressions in \cref{eq:int0_form,eq:int-1_form} are unphysical in a subtle manner.
In fact the essential discontinuity in the integrands in \cref{eq:int_norm} that are proportional to renders
their integral multi-valued. This can be seen from the result in \cref{eq:int0_form,eq:int-1_form} where the 
$z$ direction take a privileged position despite no such asymmetry in the integrand. This origin of this asymmetry 
is the order in which the integral is performed in the generalized spherical coordinates.
The multivalued integral in this case is unphysical and, as expected, puts constraints on the form of
the correlation of the noise term $\zeta$. 

One case, when the integral is single-valued corresponds to $\zeta_0 = \zeta_{-1} = 0$. 
In this case the singularity at coordinate origin is absent, and the noise is completely described
by one term:
\begin{equation}
	\avg{\zeta_i (\vb{r},t) \zeta_j (\vb{r},t)} = \zeta_1^2 \, D_{ij} \, \delta(\vb{r}-\vb{r}')\, \delta(t-t').
\end{equation}
The second case is less trivial and involves exploring the structure of \cref{eq:int0_form,eq:int-1_form}.
The integrals in these equations could be computed the same way, but with different axis choice for the spherical coordinates.
If the axis is chosen along $D_x$ or $D_y$ (instead of $D_z$ as above), 
the $D_i$-dependent pre-factors on the right hand sides of \cref{eq:int0_form,eq:int-1_form} would take a different form.
The necessary condition for the integral to be single-valued is the equality among these pre-factors 
(independent of the choice of integration variables). 
For example, if one considers the coefficient for the second term in the \cref{eq:int0_form}, one obtains the system 
of equations:
\begin{align*}
	2D_z^{-1} - D_x^{-1} - D_y^{-1} = 2D_x^{-1} - D_y^{-1} - D_z^{-1},
	\\
	2D_z^{-1} - D_x^{-1} - D_y^{-1} = 2D_y^{-1} - D_x^{-1} - D_z^{-1},
\end{align*}
whose only solution is $D_x = D_y = D_z$. When this condition is satisfied the essential discontinuity in the original
integral also vanishes, rendering the integral single-valued.
Physically this case corresponds to that of a material with isotropic thermal diffusion. Mathematically, 
this is already included in the case corresponding to $\zeta_0 = \zeta_{-1} = 0$. Thus, the latter is the only case
to be considered.

Finally, $\zeta_1$ can be computed using \cref{eq:int_for_coef}:
\begin{equation}
	\frac{k_B T^2}{C_V} = \frac{\zeta_1^2}{2V}
\end{equation}
Our result reproduces the typical results for the isotropic medium, for example~\cite{Brag99}.


\section{Origin of $\eps_{xy}'$}\label{sec:materials}

In this appendix we concern ourselves with how a nonzero $\eps'_{xy}$ can arise.

One possibility is through photoelasticity, in which applied stress produces changes in the permittivity tensor. An applied stress $\sigma_{ij}$ is linearly related to changes in the inverse dielectric tensor $B_{ij} = (\eps^{-1})_{ij}$ by $\delta B_{ij} = \pi_{ijkl} \sigma_{kl}$. To first order, perturbations in $B_{ij}$ are related to perturbations in $\eps_{ij}$ by $\delta\eps_{il} = -\eps_{ij} \delta B_{jk} \eps_{kl}$. In particular, even in a system in which the unperturbed $\eps_{ij}$ is diagonal, an off-diagonal perturbation can appear as $\delta\eps_{xy} = -\eps_{xx} \delta B_{xy} \eps_{yy}$. The way in which stresses can produce a nonzero $\delta B_{xy}$ depends on the particular crystal structure; even in cubic crystals (and isotropic materials), a shear strain $\sigma_{xy}$ will produce a nonzero $\delta B_{xy}$ via a nonzero $\pi_{xyxy}$ (in Voigt notation, $\pi_{66}$, which for cubic and isotropic materials is identical to $\pi_{44}$)~\cite{BornWolf}. A temperature-dependent term $\eps'_{xy}$ can then arise either via a temperature dependence of $\pi_{xyxy}$ or of $\sigma_{xy}$~\cite{2020JSSS....9..209S}.


\section{Field evolution with small perturbations}
\label{sec:EMevolution}

In this appendix we provide some details of the passage from the equations for
the electromagnetic field [\cref{eq:dEexpansion}] to equations for the polarization
fluctuations [\cref{eq:motion_xy}].

To derive the necessary equations we will apply several assumptions about the configuration of
electromagnetic field (notation from \cref{eq:avgE,eq:dEexpansion}):
\begin{itemize}
	\item We assume that the transverse spatial mode $f_0$ of the incident field is gaussian and that its width is much
		greater than the wavelength and temperature fluctuations scale (this is typically true for macroscopic
		mirrors, as estimated in Ref.~\cite{Brag00}). This provides us several estimates for derivatives of
		the main mode: $|\nabla f_0| \ll |kf_0|$, $|\nabla \delta e_i^{(0)}| \ll |k \delta e_i^{(0)}|$, 
		$|\delta e_i^{(0)} \nabla f_0| \ll |f_0 \nabla  \delta e_i^{(0)}|$.
	\item We assume that noise source frequency scale is much lower than the optical frequency. This
		gives us estimates of time derivatives of the fluctuations in the electromagnetic field: 
		$|\partial_t \delta e_i^{(\alpha)}| \ll |\omega \delta e_i^{(\alpha)}|$.
\end{itemize}

We plug the expansion \cref{eq:dEexpansion} into \cref{eq:dE} and project it onto the basis function $f_\alpha$. 
The projection of various terms in the equation are as follows:
\begin{equation} \label{eq_dt_mean_E}
	(f_\alpha| \partial_t^2 \avg{E_x} ) = \sqrt{P} e^{i(kn_xz - \omega t)}\left( - \omega^2 \delta_{0\alpha} \right) 
	\langle e_x^{(0)}\rangle \\
\end{equation}
\begin{multline} \label{eq:dE_dt}
	(f_\alpha| \partial_t^2 \delta E_x ) = \sqrt{P} e^{i(kn_xz - \omega t)} \times \\
		\times  \left( - \omega^2 \delta e_x^{(\alpha)} - 2 i \omega \, \partial_t \delta e_x^{(\alpha)} + \partial_t^2 \delta e_x^{(\alpha)} \right) 
\end{multline}
In the formula below we use notation $\Delta_{xy} = \partial_x^2 + \partial_y^2$:
\begin{multline} \label{eq:laplacian_dE}
	(f_\alpha| \Delta_{xy} \delta E_x ) = \sqrt{P} e^{i(kn_xz - \omega t)} \left[ \sum_\beta 
		(f_\alpha | \Delta_{xy} f_\beta) \delta e_x^{(\beta)} - \right. \\
		\left. - k^2 n_x^2 \delta e_x^{(\alpha)} + 2 i k n_x \left( \frac{\partial}{\partial z} \delta e_x^ {(\alpha)} \right)
			+\frac{\partial^2}{\partial z^2} \delta e_x^ {(\alpha)} \right]
\end{multline}
\begin{multline} \label{eq:grad_div_dE}
	(f_\alpha| \partial_x \nabla \cdot \delta \vb{E} ) =\sqrt{P} \sum_\beta \left[ e^{i(kn_xz - \omega t)}  \,
		(f_\alpha | \partial_x^2 f_\beta) \, \delta e_x^{(\beta)} + \right. \\
		+ e^{i(kn_yz - \omega t)} \, (f_\alpha | \partial_x \partial_y f_\beta) \, \delta e_y^{(\beta)} + \\
		\left. + e^{i(kn_zz - \omega t)} \, (f_\alpha | \partial_x f_\beta) \frac{\partial}{\partial z} \, \delta e_z^{(\beta)} \right]
\end{multline}
The most important case is $\alpha = 0$, since this is the projection with the highest overlap with noise
source. Several terms in the above sums can be neglected as follows:
\begin{itemize}
	\item The condition $|\partial_t \delta e_i^{(0)}| \ll |\omega \delta e_i^{(0)}|$ allows us to neglect
		$\partial_t^2 \delta e_x^{(0)}$ term in \cref{eq:dE_dt}.
	\item The relation $(f_\alpha | \partial_x f_\beta) = - (f_\beta | \partial_x f_\alpha)$ show that all
		terms of the form $(f_\alpha | \partial_x f_\beta)$ are of the order of magnitude $1 / r_0$ 
		($r_0$ is a laser beam radius), when $\alpha \sim 1$. Using the estimate, $|\nabla f_0| \ll |kf_0|$,
		terms like, $(f_\alpha | \Delta_{xy} | f_\beta)$ and
		$\partial_z^2 \delta e_x^{(\alpha)}$ in \cref{eq:laplacian_dE,eq:grad_div_dE} can be neglected for $\alpha = 0$.
\end{itemize}
Thus simplified, \cref{eq_dt_mean_E,eq:dE_dt,eq:laplacian_dE,eq:grad_div_dE} can be substituted into \cref{eq:dE},
and the $\alpha = 0$ term isolated. This gives the system of equations in \cref{eq:motion_xy}. 

\input{asymptotic_forms}

\end{document}

%% file: material_params.tex
\begin{table*}
    \caption{
    Parameters for crystalline materials.
    Values for silicon were taken from Refs.~\cite{Flubacher:1959,Glassbrenner:1964,Bradley:2006,Komma:2012}; values for lithium niobate were taken from Refs.~\cite{Nikogosyan:2005,Zelmon:97,Moretti:2005}; values for GaAs/AlGaAs were taken from Refs.~\cite{Adachi:1993,Afromowitz:1974,Kim:2007}.
    The heat capacity values here suffice for both constant-volume and constant-pressure situations, since these solids are only weakly compressible.
    The aluminum alloying fraction for AlGaAs was assumed to be \SI{92}{\%}.
    An asterisk indicates that the value was chosen ad-hoc.
    A dagger indicates that the tensor values have been assumed from scalar measurements.
    \label{tab:material_params}}
    \sisetup{table-align-text-post = false}
    \begin{tabular}{r c S S S S s}
    \toprule
    {\textbf{Quantity}}                 &
        {\textbf{Symbol}}               &
        {\textbf{Si}}                   &
        {\textbf{LiNbO$_\mathbf{3}$}}   &
        {\textbf{GaAs}}                 &
        {\textbf{AlGaAs}}               &
        {\textbf{Unit}} \\
    \midrule
    Temperature &
        $T$  &
        123  &
        293  &
        293  &
        293  &
        \kelvin \\
    Density &
        $\rho$  &
        2330    &
        4630    &
        5320    &
        3860    &
        \kg\ \meter^{-3} \\
    Heat capacity per unit mass  &
        $C$     &
        330     &
        640     &
        320     &
        440     &
        \joule\ {\kg^{-1}}\, {\kelvin^{-1}} \\
    \ldelim\{{3}{*}[Thermal conductivity]\hspace{-0.6em}  &
        $\kappa_{xx}$ &
        600\textsuperscript{\textdagger} &
        4.5 &
        44\textsuperscript{\textdagger}  &
        71\textsuperscript{\textdagger}  &
        \watt\ \meter^{-1}\ \kelvin^{-1} \\
       &
        $\kappa_{yy}$ &
        600\textsuperscript{\textdagger} &
        4.4 &
        44\textsuperscript{\textdagger}  &
        71\textsuperscript{\textdagger}  &
        \watt\ \meter^{-1}\ \kelvin^{-1} \\
       &
        $\kappa_{zz}$ &
        600\textsuperscript{\textdagger} &
        4.5 &
        44\textsuperscript{\textdagger}  &
        71\textsuperscript{\textdagger}  &
        \watt\ \meter^{-1}\ \kelvin^{-1} \\
    \midrule
    Laser wavelength in vacuum    &
        $\lambda$   &
        1550        &
        1550        &
        1550        &
        1550        &
        \nm         \\
    \ldelim\{{2}{*}[Refractive indices]\hspace{-0.6em}  &
        $n_x$    &
        3.46     &
        2.14     &
        3.37     &
        2.90     &
        -        \\
       &
        $n_y$    &
        3.46     &
        2.21     &
        3.37     &
        2.90     &
        -        \\
    \ldelim\{{3}{*}[Thermorefractive coefficients]\hspace{-0.6em}  &
        $\eps_{xx}'$      &
        700               &
        130               &
        1370              &
        1020              &
        ppm\ \kelvin^{-1} \\
       &
        $\eps_{yy}'$      &
        700               &
        -1                &
        1370              &
        1020              &
        ppm\ \kelvin^{-1} \\
       &
        $\eps_{xy}'$      &
        10\textsuperscript{*}               &
        1\textsuperscript{*}                &
        10\textsuperscript{*}               &
        10\textsuperscript{*}               &
        ppm\ \kelvin^{-1} \\
    \bottomrule
    \end{tabular}
\end{table*}

%% file: asymptotic_forms.tex

\section{Limiting forms of polarization spectral densities in transmission}\label{sec:asymptotic}

The expression for the polarization spectral densities for transmission through a crystalline material --- given 
in \cref{eq:S_xx_prefinal,eq:S_yy_prefinal,eq:S_xy_prefinal} --- can be reduced in various limiting cases to much
simpler forms. We exhibit some of these limiting cases in this appendix. Finally, in \cref{sec:adiabatic}, we 
provide an alternate calculation of $S_{e_x e_x}$ in the fully adiabatic regime, as an independent check of the
full theory in \cref{sec:anisotropic_transmission} of the main text.

\subsection{Asymptotic expansion of thermal integral}

The thermal integral [\cref{eq:Iintegral}],
\begin{equation*}
	I(\Omega_1, \Omega_2) = \int\limits_0^{\infty} \dd\tau \; \frac{\cos \Omega_1 \tau \; 
    \exp \left(-\Omega_2^2 D_{zz} \tau /c^2\right)}{ \sqrt{( \tau + \tau_x) (\tau + \tau_y)}},
\end{equation*}
dictates the frequency dependence of the polarization fluctuations for transmission through a crystalline medium.
In order to deduce limiting forms of the polarization fluctuations in the various frequency regimes of interest, it 
is germane to study the asymptotic properties of this integral.
That integral can be written as,
\begin{multline} \label{eq:final_int_slab}
	I(\Omega_1, \Omega_2) = \sum_{n = 0}^{+ \infty} C_n \left( \frac{\tau_-}{\tau_+} \right)^{2n} 
		\Re \left[e^{-i \Omega_1 \tau_+} \exp \left( \frac{\Omega_2^2}{c^2} D_{zz} \tau_+ \right) \right. \\
		\left. \times\Ei_{2n+1} \left( - i \Omega_1 \tau_+ + \frac{\Omega_2^2}{c^2} D_{zz} \tau_+ \right) \right],
\end{multline}
where $\tau_\pm = \abs{\tau_x \pm \tau_y}/2$, $C_n$ are 
the coefficients in the Taylor series expansion of function $1 / \sqrt{1 - x}$ around $x = 0$, and $\Ei_n$ stand for
the $n^\t{th}$ order exponential integral defined by, $\Ei_m(z) = \int_1^\infty \dd{t} \, t^{-m} e^{-zt}$, 
for $\Re{z} \geq 0$ and $m > 0$.
(Note that the above expansion for the integral $I$ reduces to the result obtained previously for the special case of an isotropic medium~\cite{Braginsky:2004pd}\,---\,i.e., all terms vanish except for $n=0$.) 
The physically interesting cases correspond to the argument of $\Ei_n$ approaching zero or infinity. 
The required asymptotic expansions are known for $\Ei_1$~\cite[\S2.3]{lebedev1972special}:
\begin{equation}\label{eq:ei_exand}
    \Ei_1(z) \approx 
    \begin{cases}
    \displaystyle
    - \ln z - \gamma - \sum_{ k = 1}^{\infty} \frac{(-1)^k}{k!} \frac{z^k}{k};& z \rightarrow 0 \\
    \displaystyle
    \frac{e^{-z}}{z} \sum_{ k = 0}^{N} \frac{(-1)^k k!}{z^k} + O(\abs{z}^{-N}); & z \rightarrow \infty
    \end{cases}
\end{equation}
where $\gamma \approx 0.57721$ is Euler's constant, and these are valid respectively for $\arg z \neq \pi$, and
$\arg z < \pi/2$.
Expansions for $\text{Ei}_{n>1}$ can be computed from these via the recursion relation~\cite{lebedev1972special},
$\Ei_{n+1}(z) = \int_z^\infty \dd{z'} \, \Ei_n (z')$. This relation implies that
$\text{Ei}_n(z) = o\left( \text{Ei}_1(z) \right)$ for $n>1$ when $z \rightarrow 0$,
and $\text{Ei}_n(z) \sim \text{Ei}_1(z)$ when $z \rightarrow + \infty$ 
\footnote{We use asymptotic notation in the following sense: $f(x) = O(g(x))$ when $x \rightarrow a$
means that $\lim_{x \rightarrow a} |f/g| < \infty$; $f(x) = o(g(x))$ when $x \rightarrow a$
means that $\lim_{x \rightarrow a} |f/g| = 0$; and $f(x) \sim g(x)$ when $x \rightarrow a$
means that $\lim_{x \rightarrow a} |f/g| = 1$}. 
Therefore,
only the first term in the series expansion for $I$ in \cref{eq:final_int_slab} contributes for small arguments, and all
the terms contribute the same amount for large arguments. Therefore both of the cases are solely described by the corresponding asymptotic expression for $\Ei_1$.

The above asymptotic expansions, handled carefully respecting the domain of the complex argument of $\Ei_n$ in
\cref{eq:final_int_slab}, produces the following limiting cases.

\subsection{Limiting forms of $S_{e_x e_x}$}

There are two physically interesting regimes for $S_{e_x e_x}$. 
The first regime is the small frequency limit, $n_x^2 D_{zz} \tau_+ \Omega^2 / c^2 \ll \Omega \tau_+ \ll 1$.
In this regime the power spectral density shows logarithmic behavior:
\begin{equation}
	S_{e_x e_x}(\Omega) = - \dfrac{k^2 \zeta^2 |\eps_{xx}'|^2 \ell}{16 \pi n_x^2 \sqrt{D_{yy} D_{xx}}}  
	\ln \left( |\Omega \, \tau_+| \right)
\end{equation}
The second physically interesting regime typically corresponds to the situation
when $\Omega$ is big enough that $1 \ll \Omega \tau_+$, but still small enough
for the noise to be quasistatic, i.e. $n_x^2 D_{zz} \tau_+ \Omega^2 / c^2 \ll \Omega \tau_+$.
This results in the following behavior. In this regime the power spectral density shows inverse square
behavior:
\begin{equation}
	S_{e_x e_x}(\Omega) = \dfrac{k^2 \zeta^2 |\eps_{xx}'|^2 \ell \tau_+}{8 \pi n_x^2 r_0^2}
		   \frac{1}{|\Omega \, \tau_+|^2}
\end{equation}

\subsection{Limiting forms of $S_{e_y e_y}$ and $S_{e_x e_y}$}

In addition to other parameters, power spectral densities $S_{e_x e_y}$ and $S_{e_y e_y}$ acquire additional
frequency-like parameter $\omega \Delta n$, where $\Delta n = n_x - n_y$. This results in three
physically interesting regimes. The first one corresponds to the small $\Omega$ limit when
$\Omega \tau_+ \ll D_{zz} \tau_+ \omega^2 (\Delta n)^2 / c^2 \ll 1$. In this regime the power spectral
density is frequency-independent:
\begin{equation}
	S_{e_y e_y}(\Omega) = - \dfrac{k^2 \zeta^2 |\eps_{xy}'|^2 \ell}{16 \pi n_y^2 \sqrt{D_{yy} D_{xx}}} 
		\ln \left[ \left| D_{zz} \tau_+ (\Delta n)^2 \frac{\omega^2}{c^2} \right| \right],
\end{equation}
\begin{multline}
	S_{e_x e_y}(\Omega) = \dfrac{k^2 \zeta^2 \eps_{xy}' \eps_{xx}' \ell}{16 \pi n_x n_y \sqrt{D_{yy} D_{xx}}} 
		\ln \left[ \left| D_{zz} \tau_+ (\Delta n)^2 \frac{\omega^2}{c^2} \right| \right].
	\\
	\times \frac{1 - \exp \left[-i \omega \ell \, \Delta n/c \right]}
		{\omega \ell \Delta n / c}
\end{multline}
The next regime corresponds to the transient $\Omega$ when
$D_{zz} \tau_+ \omega^2 (\Delta n)^2 / c^2 \ll \Omega \tau_+ \ll 1$.
In this regime the power spectral density shows logarithmic behavior:
\begin{equation}
	S_{e_y e_y}(\Omega) = - \dfrac{k^2 \zeta^2 |\eps_{xy}'|^2 \ell}{16 \pi n_y^2 \sqrt{D_{yy} D_{xx}}} 
		\ln \left( |\Omega \, \tau_+| \right),
\end{equation}
\begin{multline}
	S_{e_x e_y}(\Omega) = \dfrac{k^2 \zeta^2 \eps_{xy}' \eps_{xx}' \ell}{16 \pi n_x n_y \sqrt{D_{yy} D_{xx}}} 
		\ln \left( |\Omega \, \tau_+| \right).
	\\
	\times \frac{1 - \exp \left[-i \omega \ell \, \Delta n/c \right]}
		{\omega \ell \Delta n / c}
\end{multline}

Finally, there is the large $\Omega$ limit when
$D_{zz} \tau_+ \omega^2 (\Delta n)^2 / c^2 \ll \Omega \tau_+$ and $1 \ll \Omega \tau_+$.
In this regime the power spectral density shows inverse square behavior:
\begin{equation}
	S_{e_y e_y}(\Omega) = \dfrac{k^2 \zeta^2 |\eps_{xy}'|^2 \ell \tau_+}{8 \pi n_y^2 r_0^2}
		   \frac{1}{|\Omega \, \tau_+|^2}
\end{equation}
\begin{multline}
	S_{e_x e_y}(\Omega) = -\dfrac{k^2 \zeta^2 \eps_{xy}'\eps_{xx}' \ell \tau_+}{8 \pi n_x n_y r_0^2}
		   \frac{1}{|\Omega \, \tau_+|^2}
		\\
		\times \frac{1 - \exp \left[-i \omega \ell \, \Delta n/c \right]}
		{\omega \ell \Delta n / c}
\end{multline}

\subsection{Adiabatic limit: modal method}\label{sec:adiabatic}

In many situations of interest, the characteristic size of the crystal is such
that the temperature field is effectively static compared to the travel time of
the light through the crystal ($\Omega \ell \ll c$), and the fluctuations are
slow compared to the cycle of the carrier ($\Omega \ll \omega$). 
This ``adiabatic'' regime is the one considered by \citet{Braginsky:2004pd} for the
case of an amorphous material.

Here we generalize their method to the anisotropic case, with the aim of reproducing the predictions of our detailed model
through an alternate route.
The general strategy is to first compute the fluctuation in average temperature of the crystal volume probed by the laser beam, and 
then to propagate this to fluctuation in the polarization state of the light.

To find the average temperature, we start with the heat equation [\cref{eq:diffusionT}],
\begin{equation}\label{eq:adiabatic:equ}
    (\partial_t - D_{ij} \partial_i \partial_j) u(\mathbf{r},t) = \eta(\mathbf{r},t),
\end{equation}
where $\eta$ is a random heat injection with the correlation [\cref{eq:corr_T}]
\begin{equation}
    \langle \eta(\mathbf{r},t)\eta(\mathbf{r}',t')\rangle
        = \zeta^2 D_{ij} \partial_i \partial_j \delta(\mathbf{r}-\mathbf{r}') \delta(t-t')
\end{equation}
with $\zeta^2 = 2 k_\text{B} T^2 / c_V$.

\begin{widetext}
Given the boundary condition $\partial_z u = 0$ at $z=0$ and $z=\ell$, we can write down a series solution
\begin{equation}\label{eq:adiabatic:urt}
    u(\mathbf{r},t) = \int\limits_{-\infty}^\infty \frac{\rmd k_x\,\rmd k_y\,\rmd\Omega}{(2\pi)^3} \sum_n u_n(k_x,k_y,\Omega) \rme^{\rmi\Omega t - i k_x x - i k_y y} \cos(b_n z)
\end{equation}
with $b_n = \pi n / \ell$. Each coefficient,
\begin{equation}
    u_n(k_x, k_y, \Omega) = \int\limits_{-\infty}^\infty \rmd x \, \rmd y \, \rmd t \rme^{-\rmi\Omega t + \rmi k_x x + \rmi k_y y} \int\limits_0^\ell \rmd z \, \frac{2 - \delta_{0n}}{\ell} \cos(b_n z) u(x, y, z, t),
\end{equation}
defines a mode of the local temperature field.
Inserting the above expansion [\cref{eq:adiabatic:urt}] in the heat equation [\cref{eq:adiabatic:equ}], it is found that
each mode is independent, and given by,
\begin{equation}
    u_n(k_x, k_y, \Omega) = \frac{\eta_n(k_x, k_y, \Omega)}{\rmi\Omega - D_{ij} k_i k_j},
\end{equation}
where,
\begin{equation}
    \langle \eta_m(k_x,k_y,\Omega) \eta_n^*(k_x', k_y', \Omega') \rangle
        = (2\pi)^3 \zeta^2 D_{ij} k_i k_j \frac{2-\delta_{0n}}{\ell} \delta_{mn}
            \delta(k_x - k_x') \delta(k_y - k_y') \delta(\Omega - \Omega').
\end{equation}
We now construct an ad-hoc observable~\cite{Levin98,Levin08}, 
the volume-averaged temperature over the cylindrical region of the beam (radius $r_0$) in the crystal (length $\ell$):
\begin{equation}
    \bar{u}(t) = \frac{1}{\pi r_0^2 \ell} \int\limits_{V}\! \rmd^3\mathbf{r}\, \rme^{-(x^2+y^2)/r_0^2} u(\mathbf{r},t) 
    = \int\limits_{-\infty}^{+\infty} \frac{\rmd k_x \,\rmd k_y\,\rmd\Omega}{(2\pi)^3} \rme^{-r_0^2 (k_x^2 + k_y^2) / 4} 
    \rme^{\rmi\Omega t} u_0(k_x, k_y, \Omega);
\end{equation}
where we have dropped all terms in the Fourier series except for the $n=0$ term, since this is the only term which has a non-zero integral in 
the $z$ direction.
We can compute the correlation function of this volume-averaged temperature, giving
\begin{equation}
    \langle \bar{u}(t)\bar{u}(t+\tau) \rangle
        = \zeta^2 \times \frac{1}{\ell} \int\limits_{-\infty}^{+\infty} \frac{\rmd k_x \,\rmd k_y\,\rmd\Omega}{(2\pi)^3} \rme^{-r_0^2 (k_x^2+k_y^2)/2} \frac{\mathfrak{D}(k_x,k_y) \,\rme^{\rmi\Omega\tau}}{\Omega^2 + \mathfrak{D}(k_x,k_y)^2},
\end{equation}
where $\mathfrak{D}(k_x,k_y) = D_{xx} k_x^2 + 2D_{xy} k_x k_y + D_{yy} k_y^2$.
Then since the correlation function is related to the two-sided spectral density $\mathcal{S}(\Omega)$ by $\langle\overline{u}(t)\overline{u}(t+\tau)\rangle = \tfrac{1}{2\pi} \int_{-\infty}^{+\infty} \dd{\Omega} \, \mathcal{S}(\Omega) \rme^{\rmi\Omega\tau}$, we can immediately read off the one-sided spectral density $S(|\Omega|) = 2\mathcal{S}(\Omega)$ for the volume-averaged temperature:
\begin{equation} \label{eq:BGV_Suu}
    S_{\bar{u} \bar{u}}(\Omega) = \frac{4 k_\text{B} T^2}{c_V \ell} \int\limits_{-\infty}^{+\infty} \frac{\rmd k_x \, \rmd k_y}{(2\pi)^2} \frac{\mathfrak{D}(k_x,k_y)\, \rme^{-r_0^2 (k_x^2 + k_y^2)/2}}{\Omega^2 + \mathfrak{D}(k_x,k_y)^2},
\end{equation}
which reduces to Eq.~(E6) of \citet{Braginsky:2004pd} in the thermally isotropic limit. 
Note that we did not use any assumption about the length scale of the crystal relative to the thermal diffusion wavelength in this derivation.
\end{widetext}

To complete the calculation, we need to connect fluctuations in $\bar{u}(t)$ to fluctuations in the optical field. The variable 
$\bar{u}(t)$ has been constructed so that in an optically isotropic material, the phase fluctuation of a passing beam is computed by
\begin{equation}
    \delta\phi(t) = (\omega/c) \ell \beta \bar{u}(t)
\end{equation}
with $\beta = \partial n/\partial T$.
This equation is equivalent to \cref{eq:ex_int} in the limit $\ell \Omega \ll c$, in which case the retardation term $n_i(z-z')/c$
in \cref{eq:ex_int} can be omitted (i.e., we neglect the light travel time through the crystal).
This approximation holds even for meter-scale optics so long as $\Omega/2\pi\lesssim\SI{50}{\MHz}$. 
Thus we arrive at the limiting form,
\begin{equation}\label{eq:Sxx_adiabatic}
    S_{e_x e_x}(\Omega) = \left|\frac{k\ell\eps'_{xx}}{2n_x} \langle e_x^{(0)}\rangle \right|^2 S_{\bar{u}\bar{u}}(\Omega),
\end{equation}
valid in the adiabatic regime.
Note that we cannot arrive at a similar limiting expression for $S_{e_y e_y}$ or $S_{e_x e_y}$ using the modal expansion method 
because the assumption that only the $n=0$ mode contributes to $\bar{u}$ no longer holds.

\subsection{Adiabatic limit: direct method}\label{sec:levin}

We now perform a second independent check of our formalism by modeling the transmission problem 
using Levin's approach via the fluctuation-dissipation theorem \cite{Levin98,Levin08}.

Levin directly computes the spectral density of an ad-hoc ``observable'',
\begin{equation}
	\int \dd{V} \; q(\vb{r}) \, \delta u (\vb{r}, t),
\end{equation}
whose form is intuited to reflect the transduction of local temperature fluctuations to the relevant
optical property. We take $\delta e_x$ to be the observable of interest, in which case,
\begin{equation}
	q(\vb{r}) = \frac{k \eps'_{xx} \langle e_x^{(0)}\rangle}{2\pi n_x r_0^2} 
		\exp \left[ - \frac{x^2 + y^2}{r_0^2} \right].
\end{equation}
This is read off from \cref{eq:ex_int}.

Next, Levin studies how a sinusoidal injection of entropy,
\begin{equation}
	\frac{\delta s}{\delta V} = F_0 q(\vb{r}) \cos \Omega t,
\end{equation}
is distributed in the medium via thermal dissipation. This can be done in the anisotropic case by solving
the sinusoidally-driven heat equation,
\begin{equation}
	(\partial_t - D_{ij} \partial_i \partial_j) \delta T 
	= \frac{T}{c_V} \frac{\partial}{\partial t} \left( \frac{\delta s}{\delta V} \right),
\end{equation}
with insulating boundary conditions at $z=0,\ell$.
The required solution is,
\begin{multline}
	\delta T = \frac{k \eps'_{xx} \langle e_x^{(0)}\rangle F_0 \Omega}{4 i n_x c_V} \int\limits_{- \infty}^{+ \infty} 
	\frac{\dd{k_x}\, \dd{k_y}}{(2 \pi)^2} \times
		\\
		\left[ \frac{e^{i (\Omega t - k_x x - k_y y)} e^{- (k_x^2 + k_y^2) r_0^2/4}}{i \Omega + \mathfrak{D}(k_x,k_y)} - \text{c.c.}\right].
\end{multline}

It is in this step that our approach diverges from that of Levin's. We solve for the stochastic local temperature
field by augmenting the thermal transport equation with a source that is consistent with the known equilibrium
temperature fluctuation (essentially a fluctuation-dissipation theorem for the temperature). We then propagate
that local temperature field through its impact on the electromagnetic field. Levin directly applies the
fluctuation-dissipation theorem to the ad-hoc observable. 

To do so, it is necessary to compute the dissipated energy. In the anisotropic case, it is,
\begin{equation}
	W_{\t{diss}} = \int \dd{V} \; \frac{\kappa_{ij}}{T} \langle \partial_i(\delta T) \partial_j(\delta T) \rangle.
\end{equation}
Knowing the dissipated energy, the fluctuation-dissipation theorem can be applied to derive the spectral density
of the observable,
\begin{equation}
	S_{e_x e_x} = \frac{8 k_B T}{\Omega^2} \frac{W_{\t{diss}}}{F_0^2} 
	= \left|\frac{k\ell\eps'_{xx}}{2n_x} \langle e_x^{(0)}\rangle \right|^2 S_{\bar{u}\bar{u}}(\Omega),
\end{equation}
where $S_{\bar u \bar u}$ is given by \cref{eq:BGV_Suu}.
As expected, Levin's method and the first principles calculations agree in the adiabatic limit.